\documentclass[useAMS,usenatbib,usegraphicx]{mn2e}
\newcommand{\aap} {{\it A\&A}}
\newcommand{\aj} {{\it AJ}}
\newcommand{\apj} {{\it ApJ}}
\newcommand{\apjl} {{\it ApJL}}
\newcommand{\apjs} {{\it ApJS}}
\newcommand{\araa} {{\it ARA\&A}}
\newcommand{\mnras} {{\it MNRAS}}

\newcommand{\hii} {H{\sc ii}}
\newcommand{\kms}{{\rm ~km~s}^{-1}}

\newcommand{\Mg} {M_{\rm g}}

\newcommand{\mst}{m_{*}}
\newcommand{\msp}{m_{\rm SP}}
\newcommand{\Msun}{{\rm M}_{\odot}}

\newcommand{\Ng} {N_{\rm g}}
\newcommand{\nlos}{n_{\rm LOS}}
\newcommand{\pcc} {{\rm cm}^{-3}}
\newcommand{\Rg} {R_{\rm g}}
\newcommand{\tage} {t_{\rm age}}

\def\cm3{\mbox{cm$^{-3}$}}
\def\msun{\mbox{M$_\odot$}}

\def\nsf{\mbox{$n_{\rm SF}$}}
\def\Ke{\mbox{K}}

\title[Hierarchical Cluster Assembly in Collapsing Clouds]
{Hierarchical Cluster Assembly in Globally Collapsing
Clouds}
\author[V\'azquez-Semadeni
  et al.]{Enrique V\'azquez-Semadeni$^{1}$\thanks{ e-mail:
      e.vazquez@crya.unam.mx}, Alejandro Gonz\'alez-Samaniego$^{2}$, and
Pedro Col\'in$^{1}$\\
$^{1}$Instituto de Radioastronom\'ia y Astrof\'isica, Universidad
  Nacional Aut\'onoma de M\'exico, Apdo. Postal 3-72, Morelia, 58089, M\'exico \\
$^{2}$ Center for Cosmology, Department of Physics and Astronomy, University of California at Irvine,
Irvine, CA 92697, USA}

\begin{document}

\maketitle
\label{firstpage}

\begin{abstract}
  We discuss the mechanism of cluster formation in a numerical
  simulation of a molecular cloud (MC) undergoing global hierarchical
  collapse (GHC), to understand how the gas motions in the parent
  cloud control the assembly of the cluster. The global nature of the
  collapse implies that the star formation rate (SFR) increases over
  time. The ``hierarchical'' nature of the collapse consists of
  small-scale collapses within larger-scale ones. The large-scale
  collapses culminate a few Myr later than the small-scale collapses and
  consist of filamentary flows that accrete onto massive, dense central
  clumps. In turn, the small-scale collapses form clumps embedded in the
  filaments, that are falling onto the central clump assembled by the
  ongoing large-scale collapse. The stars formed in the early,
  small-scale collapses share the infall motion of their parent clumps
  towards the larger-scale potential well, so that the filaments feed
  both gaseous and stellar material to the massive central clump. This
  leads to the presence of a few older stars in a region where new
  protostars are forming at a higher rate, and also to a self-similar or
  fractal-like structure of the clusters, in which each unit is composed
  of smaller-scale sub-units, which approach each other and may
  eventually merge, explaining the frequently-observed morphology of
  cluster-forming regions. Moreover, because the older stars formed in
  the filaments share the infall motion of the gas onto the central
  clump, they tend to have larger velocities and to be distributed over
  larger areas than the younger stars formed in the central clump, where
  the gas from which they form has been shocked and has dissipated some
  kinetic energy. Finally, interpreting the IMF as a probability
  distribution, so that the probability of forming a massive star is
  much lower than that of forming a low-mass one, implies that massive
  stars only form once the {\it local} SFR is large enough to sample the
  IMF up to high masses. In combinaton with the increase of the SFR,
  this implies that massive stars tend to appear late in the evolution
  of the MC, and only in the central massive clumps. We discuss the
  correspondence of these features with observed properties of young
  stellar clusters, finding very good qualitative agreement, thus
  providing support to the scenario of global, hierarchical collapse of
  MCs, while explaining the origin of the observed cluster structure.
\end{abstract}


\section{Introduction} \label{sec:intro}

It is presently accepted that most stars form in clusters or groups,
although the details of the cluster-formation process, especially the
origin of their structural properties, remain a matter of active
research \cite[see, e.g., the reviews by][]{LL03, PZ+10, Moraux16}. In
particular, in recent years, a number of structural properties of the
clusters have been uncovered that still require an adequate theoretical
understanding, such as: i) The existence of a mass segregation in the
clusters, with the most massive stars lying closer to the cluster's
center \cite[]{HH98}; ii) the distribution of protostellar separations,
which appears to have no characteristic scale \cite[]{Bressert+10}; iii)
the likely existence of an age gradient in clusters, with the youngest
stars being located in the highest-density regions \cite[]{Kuhn+15a};
iv) the apparent deficit of OB stars in some infrared dark clouds
\citep[IRDCs;] [] {Povich+16}.

Numerical simulations have begun to offer some insight about these
properties. \cite{Kirk+14} have concluded, from a suite of simulations
of self-gravitating, decaying isothermal turbulence, that the most
massive stars form in situ at the cluster centers, rather than
``sinking'' there through dynamical interactions in the cluster
itself. However, they gave no physical explanation as to why the most
massive stars should form there. More recently, \cite{Kuhn+15b} have
suggested, by comparing multi-wavelength observations of stellar
clusters with numerical simulations, that clusters form by mergers of
``subcluster'' structures, although again no explanation of why such
mergers should occur is provided \citep[see also] [for N-body studies of
subcluster merging] {McMillan+07, Allison+09, MB09}. Observationally,
the presence of subunits of somewhat different ages in the clusters has
also been pointed out by \cite{RG+15}.

A physical mechanism capable of providing a unifying scenario to these
properties is that of global, hierarchical molecular cloud collapse,
advanced by \citet[] [see also G\'omez \& V\'azquez-Semadeni 2014]
{VS+09}. The latter authors noted that, if molecular clouds (MCs) are
assembled by large-scale colliding streams of warm, atomic gas that
rapidly condenses into the cold atomic phase, then they quickly become
Jeans-unstable and begin to collapse globally. Moreover, the turbulence
induced by the collision of the streams causes moderately supersonic
turbulence \cite[e.g.,][]{KI02, Heitsch+05} in the cold gas, which
produces a multi-scale spectrum of density fluctuations, where 
small-scale, large-amplitude (SSLA) density fluctuations are superposed on
larger-scale, smaller-amplitude (LSSA) ones \cite[e.g.,][] {KR05}. Since
these density fluctuations are nonlinear, the denser SSLA fluctuations have
shorter free-fall times than the LSSA ones, therefore completing their
collapse earlier.  This process is therefore similar to \cite{Hoyle53}
fragmentation, except that the density fluctuations are of turbulent
origin and therefore nonlinear. In this sense, the process is also
similar to the mechanism of ``gravo-turbulent fragmentation''
\citep[e.g.,] [and references therein] {MK04, BP+07}, except that the
cloud is {\it not} globally supported by turbulence, and the turbulent
fluctuations do not collapse directly, but rather just plant the seeds
for subsequent, local, scattered collapses as the cloud contracts
globally \citep{CB05}.  In what follows, we will refer to this
intermediate scenario between Hoyle and gravoturbulent fragmentation as
``global hierarchical collapse'' (GHC).

This scenario also predicts that the star formation rate (SFR) in MCs
evolves (initially increasing) over time, as a consequence of the
increase of the mean density of the clouds as they go through global
gravitational collapse \citep{ZA+12, Hartmann+12, ZV14}. Assuming that
massive stars do not form until the SFR is high enough that the IMF is
sampled up to high masses, then massive stars form late in the global
process, and when they do, they begin to disrupt their parent clouds
through their feedcak (winds, ionising radiation, SN explosions),
reducing the SFR again \citep[e.g., ] [] {VS+10, ZA+12, Colin+13, ZV14,
Dale+14, Dale+15, Gatto+16}.

\citet{GV14} presented a smoothed-particle hydrodynamics (SPH) numerical
simulation of cloud formation and evolution in the context of GHC that
showed the formation of filamentary structures with embedded clumps. In
that simulation, the filaments constitute river-like structures, through
which material flows from the extended cloud environment to the dense
cores where star formation occurs. When the filaments are sufficiently
dense, fragmentation occurs within them as the gas flows along them into
the most massive cores. This implies that the filaments supply the dense
cores with a mixture of stars and gas.

This kind of flow was also observed (although it was not discussed) in a
similar simulation presented by \citet[] [herefater Paper I] {Colin+13}
using the adaptive mesh refinement code ART \citep{KKK97, Kravtsov+03}
that included a simplified treatment of radiative transfer and a
prescription to form stellar particles\footnote{We refrain from calling
  the stellar particles ``sinks'' because, contrary to the standard
  practice for sink particles, our stellar particles are not allowed to
  accrete. Instead, in our prescription, the accretion is done in the
  gas phase (cf.\ Sec \ref{sec:SF}).} (SPs) that allows imposing a
power-law SP mass function with a slope similar to that of
\citet{Salpeter55}. This implies that, contrary to the situation in the
simulation by \citet{GV14}, the clusters formed in the simulation of
Paper I, as well as their surrounding gas, are subject to realistic
dynamics, which allows investigating the evolution of the clusters from
their formation to the time when they disperse their surrounding gas.

There exist many numerical studies of cluster formation, focusing on
issues such as their stellar mass function, the correlation function of
the spatial stellar distribution and cluster boundedness, and the
formation of binaries \citep[e.g.,] [] {KB01, BBV03, Bate09a}; the
effect of feedback on producing massive stars \citep[e.g.,] []
{Bate09b, Krumholz+10, Krumholz+12}
and on destroying their parent clumps \citep[e.g.,] [Paper I]
{VS+10, Dale+12, Dale+13a, Dale+13b}; and the energy balance and
rotation of the cluster as a function of the initial turbulence level in
the parent cloud \citep{LH16}. 


In this work, instead, we aim to describe the process of assembly and
early evolution of the clusters as a consequence of GHC. To this end, we
study a cluster formed in the simulation labeled LAF1 in Paper I,
focusing on the resulting spatial structure of the cluster. In Sec.\
\ref{sec:model} we briefly describe the numerical simulation, and in
Sec.\ \ref{sec:cluster} we describe the criteria for defining the
cluster and stellar groups, both in terms of the origin of their members
as well as from their instantaneous positions. Next, in Sec.\
\ref{sec:results} we present our results concerning the assembly from
subunits brought in by the GHC, as well as the resulting structure of
the clusters. In Sec.\ \ref{sec:discussion} we discuss the implications
of our results and compare them with existing observations, and in Sec.\
\ref{sec:conclusions} we give a summary and some conclusions.

\section{The Numerical Model} 
\label{sec:model}

The numerical simulation used in this work comes from the set performed
in Paper I with the hydrodynamics+N-body Adaptive Refinement Tree (ART)
code of \cite{Kravtsov+03}. The physical processes included are
self-gravity, parameterized heating and cooling (Sec.\
\ref{sec:HCrates}), star formation with an imposed but realistic IMF (Sec.\
\ref{sec:SF}), and simplified radiative transfer for the feedback from
massive-star ionising radiation (Sec.\ \ref{sec:feedback}).  Magnetic
fields are neglected. For more details, we refer the reader to Paper I.

\subsection{The numerical box and resolution} \label{sec:box_res}

The simulation represents the head-on collision of two cylindrical
streams in the warm neutral medium. The streams each have a radius of 64
pc and a length 112 pc, and each is traveling at a speed of $5.9\,
\kms$. The numerical box (including the streams) initially has a uniform
density $n = 1\, \pcc$ and temperature $T = 5000$ K, implying an
adiabatic sound speed of $7.4\, \kms$. Thus, the streams move with a
Mach number of 0.8 with respect to the sound speed in the initial
uniform background medium.

The simulation uses a base resolution of $128^3$ grid cells, and allows
for 5 refinement levels, reaching a maximum resolution equivalent to
$4096^3$, with a minimum cell size of 0.0625 pc, or $\approx $ 13 000
AU. The refinement is based on a ``constant mass'' criterion, so that a
cell size is refined when its mass exceeds $0.32 \Msun$. This implies
that the grid cell size $\Delta x$ scales with density $n$ as $\Delta
x \propto n^{-1/3}$. Once the maximum refinement level is reached, no
further refinement is performed, and the cell's mass can reach much
larger values because of the probabilistic SF scheme (Sec.\ \ref{sec:SF}).

Note that this constant-cell-mass refinement criterion does not conform
to the so-called {\it Jeans criterion} \citep{Truelove+97} of resolving
the Jeans length with at least 4 grid cells. Those authors
cautioned that failure to do this might result in spurious, numerical
fragmentation. However, we do not consider this a cause for concern
since, as will be described in Sec.\ \ref{sec:SF}, our star
formation prescription allows us to choose the stellar-particle mass
distribution, and tune it to a \citet{Salpeter55} value.

\subsection{Heating and Cooling} \label{sec:HCrates} 

We use heating ($\Gamma$) and cooling ($\Lambda$) functions of the form  
\begin{eqnarray} 
\Gamma &=& 2.0 \times 10^{-26} \hbox{ erg s}^{-1}\label{eq:heating}\\
\frac{\Lambda(T)}{\Gamma} &=& 10^7 \exp\left(\frac{-1.184 \times
10^5}{T+1000}\right) \nonumber \\
&+& 1.4 \times 10^{-2} \sqrt{T} \exp\left(\frac{-92}{T}\right)
~{\rm cm}^3.\label{eq:cooling}
\end{eqnarray}
These functions are fits to the various heating and cooling processes
considered by \citet{KI00}, as given by equation (4) of \citet{KI02},
with the typographical corrections noted in \citet{VS+07}.  With these
heating and cooling laws, the gas is thermally unstable in the density
range $1 \la n \la 10\ \cm3$. No chemistry tracking is performed, and so
the gas is assumed to be all atomic, with a mean particle weight $\mu
=1.27$.

\subsection{Star formation} \label{sec:SF} 

Our simulation employs a probabilistic star formation prescription such
that, if a grid cell reaches a number density $n > \nsf$, where \nsf\ is
a density threshold, then an SP of fixed mass $\msp$
may be placed in the cell with a probability $P$ every timestep of the
root (coarsest) grid.  If the SP is created, it acquires half of the gas
mass of the parent cell.

Unlike standard sink particles \citep[e.g.,][]{Bate+1995, Jappsen+05,
  VS+07, Federrath+2010}, our SPs do not accrete.  However, since our
prescription for SF is probabilistic, once the highest refinement level
is reached in a cell, it actually continues to accrete from the
surrounding material, and grows in mass and density until an SP is
actually placed in it. This implies that the accretion is performed in
the gas phase, rather than onto the particles.  Serendipitously, it was
found in Paper I that this prescription allows the SPs to form with a
power-law mass distribution, whose slope can be tuned by varying the
value of $P$, at a given maximum resolution.  In Paper I it was reported
that setting $P = 3 \times 10^{-3}$ at the resolution used in our
simulation caused the SP power-law mass distribution to attain an
exponent of $-1.34$, thus being similar to the classical Salpeter value.
We set the threshold density for SP formation at $\nsf = 9.2 \times
10^4\ \cm3$, which corresponds to a cell mass of $0.78 \Msun$ at the
highest refinement level. Thus, the minimum possible SP mass is $\msp =
0.39 \Msun$. With this prescription, the most massive SP formed in this
simulation has $\msp =61\ \msun$.

Note, however, that our imposed IMF is a strict power law, and thus the
most probable stellar mass is also the minimum. This implies that we are
missing the low-mass ($M < 0.39 \Msun$) side of the actual IMF, implying
that we are missing about half of the total number of stars, and that
their corresponding mass (20--30\% of the total mass) is deposited
instead into the stars that do get formed in the simulation. We discuss
the implications of this in Sec.\ \ref{sec:limits}.

Our simulation thus contains SPs that represent individual stars (not
clusters or groups), with a realistic mass distribution above the most
probable stellar mass. This allows the dynamics of the SPs to be
reasonably representative of the forming cluster dynamics, contrary to
the case when the particles' masses correspond to those of small groups,
of up to a few hudreds of solar masses. In that case, low-mass
particles can suffer exceedingly strong encounters with the massive
particles, and thus acquire unrealistically high velocities.


\subsection{Feedback prescription} \label{sec:feedback}

In our simulation, the feedback effect of
the ionising stars on the MC is non-local, using a simplified radiative
transfer prescription
to model the formation and evolution of a HII region around an SP. As a
zeroth-order approximation, we assume a uniform ``line-of-sight''
(LOS) characteristic density $\nlos$ in the
HII region, with a value equal to the geometric mean between the density
of the cell containing the SP and that of the target cell. Moreover, for
any given star we assume an ionising flux $S_*$ taken from the tabulated
data provided by \citet{Diaz+98}. We then compute the corresponding
Str\"omgrem radius, $R_{\rm S}$, given by
\begin{equation}
R_{\rm S} \equiv \left( \frac{3}{4 \pi} \frac{S_*}{\alpha \nlos^2}
\right)^{1/3},
\label{eq:Rs}
\end{equation}
where $\alpha = 3.0 \times 10^{-13}$ cm$^3$ s$^{-1}$ is the hydrogen
recombination coefficient.

At each time step, we search for all cells surrounding every SP particle
whose distances to it are smaller than $R_{\rm S}$. For each target cell that
satisfies this condition, we set its temperature to $10^4\ \Ke$ and turn
off the cooling there. The cell will remain so during the whole
lifetime of the star, which is determined by its mass $\mst$ as
\begin{equation}
t_{\*} = \left\{ \begin{array}{ll}
      \mbox{2 Myr} & \mbox{if $\mst \le 8 \Msun$};\\
      222 ~\mbox{Myr}~\left(\frac{\mst}{\Msun} \right)^{-0.95} & \mbox{if $\mst > 8 \Msun$}.\end{array} \right.
\label{eq:st_lifetime}
\end{equation}
As explained in Paper I, for stars more massive than $8 \Msun$, this
time is a fit to the stellar lifetimes by \citet{Bressan+93}, while for
stars with masses lower than that, it represents the fact that the
duration of the stellar-wind phase is $\sim 2$ Myr, roughly
independently of mass. This also means that we are representing the
effect of the winds and outflows of low mass stars by an ionization
prescription. While this is clearly only an approximation, we do not
expect it to have much impact on our calculations, since the main source
of feedback energy at the level of GMCs is the ionization feedback from
massive stars \citep{Matzner02}.

Although highly simplified, this prescription was tested in Paper I
against the known analytic formula for the evolution of $R_s$ in a
uniform medium \citep{Spitzer78} and shown to agree within 30\%.  This,
we believe, is accurate enough given our interest in large-scale
molecular cloud evolution and the associated cluster dynamics at early
stages of assembly, rather than on the structural details of the ionised
regions.

\section{Cluster definition: membership} 
\label{sec:cluster}


One crucial step for the present study is the membership of stars to a
given group or cluster. We now discuss our procedure of choice.

Observationally, the identification and membership definition of
embedded stellar clusters is intrinsically a complex task, complicated
by ambiguities and uncertainties, since the clusters do not have well
defined edges, contamination by field stars is difficult to quantify,
and faint members may be difficult to identify \citep[e.g.,] [] {LL03}.
In our case, the origin of the stars in the simulations is known, and
this allows a classification of the stars by site of origin.  However,
this information is unavailable to observations, which instead must
proceed by first identifying the cluster as a stellar surface
density excess, and then define its members mainly on a statistical
basis, by comparison with star counts in nearby control fields off the
cluster \citep{LL03}. On the other hand, in the simulation there are no
``field'' stars, but only stars from clusters formed at some other
locations in the cloud.

 

For the purpose of defining a cluster or group\footnote{We use the terms
``cluster'' and ``group'' in a somewhat loose way to respectively denote
larger and smaller associations of stars, the latter usually being part
of the former. The looseness comes from the
fact that, given the hierarchical and evolutionary nature of the
process, both clusters and groups evolve and grow in stellar count, mass
and size.} by location at early times,
we use the well known friends-of-friends (FOF) algorithm
\citep{Davis+85}. This is convenient at early stages, when the first
groups 
of stars are well separated and the stars are easy to tag as part of a
group that has formed at a certain time and location. However, note that,
because of the hierarchical structure of the clusters,\footnote{In our
simulation, stars are born naturally in groups and subgroups; see Secs.\
\ref{sec:intro} and \ref{sec:results}.} the number and stellar content
of the resulting groups depends on the value of the so-called ``linking
parameter'' passed to the algorithm. This parameter is a measure of the
distance out to which neighbors are searched. At early times, the
separation between groups is easily determined by eye, and the linking
parameter can then be chosen to match the visual classification.

At later times, however, the groups change their size by dynamical
interactions between their members, and moreover new stars continue to
form in the same region, or in its outskirts, making it necessary to
suitably define the regions' size in order to determine stellar
membership to a group. We thus define the radius of a group at a given
time by performing an iterative process: 

\begin{enumerate}

\item Considering only the stars that were part of the group in the
  previous snapshot, we measure the distance from their center of mass
  to the most distant star. If this distance is less than twice the
  distance from the center to the second most distant star, we consider
  the distance to the most distant star as the radius of the group.
  Otherwise, we consider the most distant star as a ``runaway star'' and
  take the radius of the group as the distance from its center to the
  second most distant star.

\item Next, we
assign the new stars (those
formed in the period of time between the actual snapshot and the
previous one) to the group whose center
of mass is closest to them. 

\item Finally, including the new stars, we recompute the center of mass,
  size ($\Rg$), total mass ($\Mg$), and total number of stars ($\Ng$) of
  the group. We also
  compute the age of the group, $\tage$, as the time elapsed since the
  moment when the group was first defined.

\end{enumerate}

However, neighbouring groups generally approach each other (as part of
the large-scale collapse) and often they merge. We say that a new group
has formed by the merger of two former groups when the distance between
their centers of mass is smaller than the larger of the two radii. At
that point, we consider that the earlier subgroups disappear. In
general, we expect this process to operate at all scales in our
simulation, with groups in turn forming from subgroups, and so on, in a
hierarchical and self-similar manner.

\section{Results} 
\label{sec:results}

\subsection{Global evolution and hierarchical cluster assembly} \label
{sec:global_evol} 

\begin{figure*}
\begin{center}
 \includegraphics[width=0.9\textwidth]{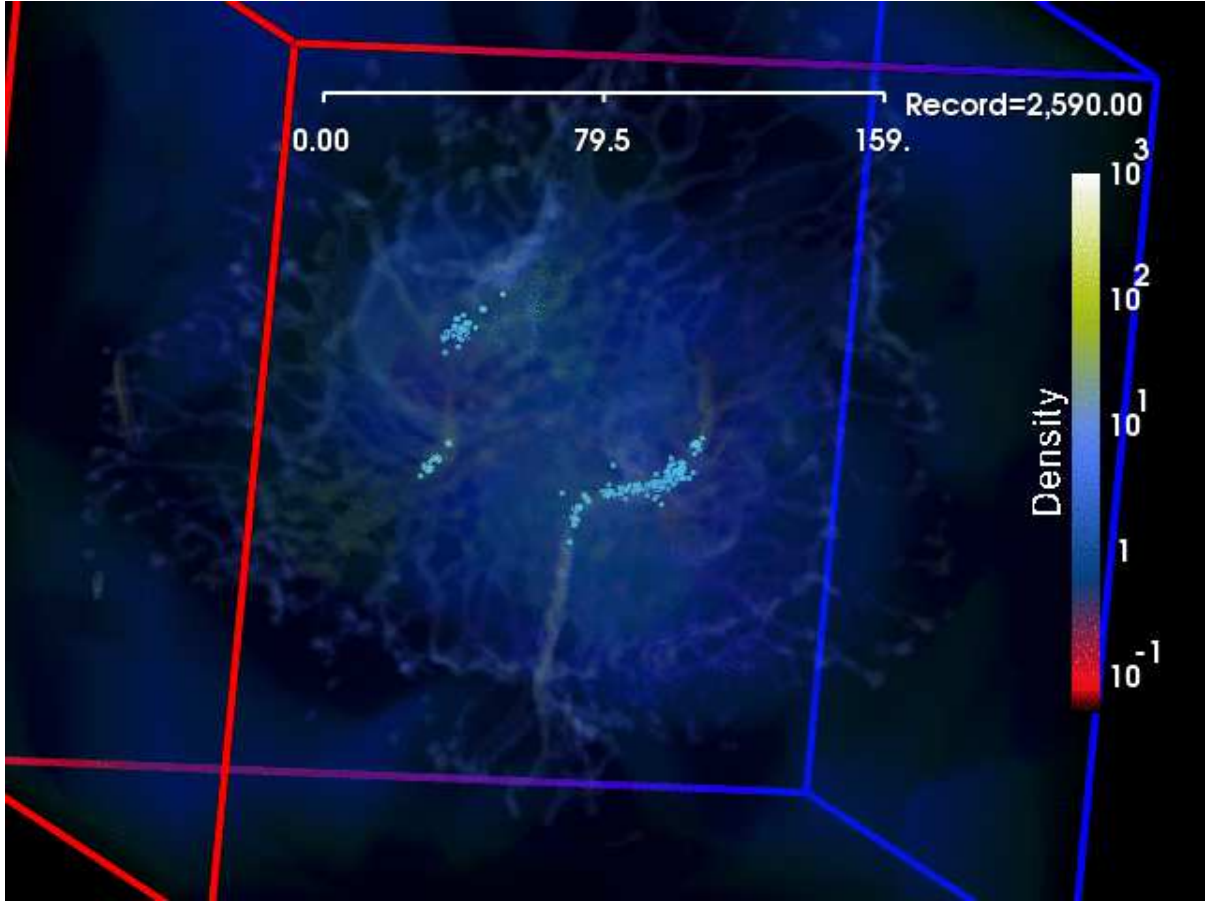} 
 \caption{A global projected view at $t=25.9$~Myr, showing the three
   clusters that form within filamentary structures. The box size is 256
   pc and the ruler shows a scale of 159 pc. The color bar indicates the
   local density in $\pcc$. The cyan dots represent individual stars.
   Expanding shells can be seen around the two most populous clusters.}
\label{fig:sim_imgs}
\end{center}
\end{figure*}

In this simulation, three main clusters form, two of which are
sufficiently massive to clear out the dense gas around them on a
timescale $\tau_{\rm clear} \sim 17$ Myr since the formation of the
first stars (from $t \sim 19$ Myr to $t \sim 36$ Myr in the evolution of
the simulation). Stars begin to form at $t
\approx 18.9$ Myr, and the first \hii\ regions appear at $t \approx
24.2$ Myr.\footnote{Here, it is important to recall that the GHC
scenario implies that the first episodes of SF involve small numbers of
stars, and thus they do not manage to sample the IMF out to large
stellar masses. The early SF episodes occur in a ``scattered'' mode,
which corresponds to the culmination of the collapse of the low-mass,
high-density, small-scale (SSLA) fluctuations that produce low numbers of
stars, and that are themselves falling into the large-scale potential
wells that later become the high-mass SF sites \citep{VS+09}.}
Specifically, at $t \approx 37.5$ Myr, the dense gas has been cleared
from a radius of $\sim 70$ pc around the clusters.  Figure
\ref{fig:sim_imgs} shows one snapshot of the
simulation, at $t=25.90$ Myr into the evolution, in which two
\hii\ regions, surrounding the two most massive clusters, can be seen, while
the filamentary structure of the cloud is still noticeable in
general. The clusters themselves have formed along the main filamentary
structures in the cloud, and contain $10^3$--$10^4
\Msun$. Note also that the structure of the most massive cluster retains
the filamentary shape of its parent cloud, similarly to, for example,
the structure observed in the Orion A cloud \citep[see, e.g., Fig.\ 1
of] [] {Hacar+16}. This similarity  in fact is not limited to the
filamentary shape, but also includes the fact that the filament has a
larger concentration of stars in one end. In a future paper we plan to
quantitatively explore the stucture and kinematics of these observed and
simulated clusters.

\begin{figure*}
\begin{center}
 \includegraphics[width=0.49\textwidth]{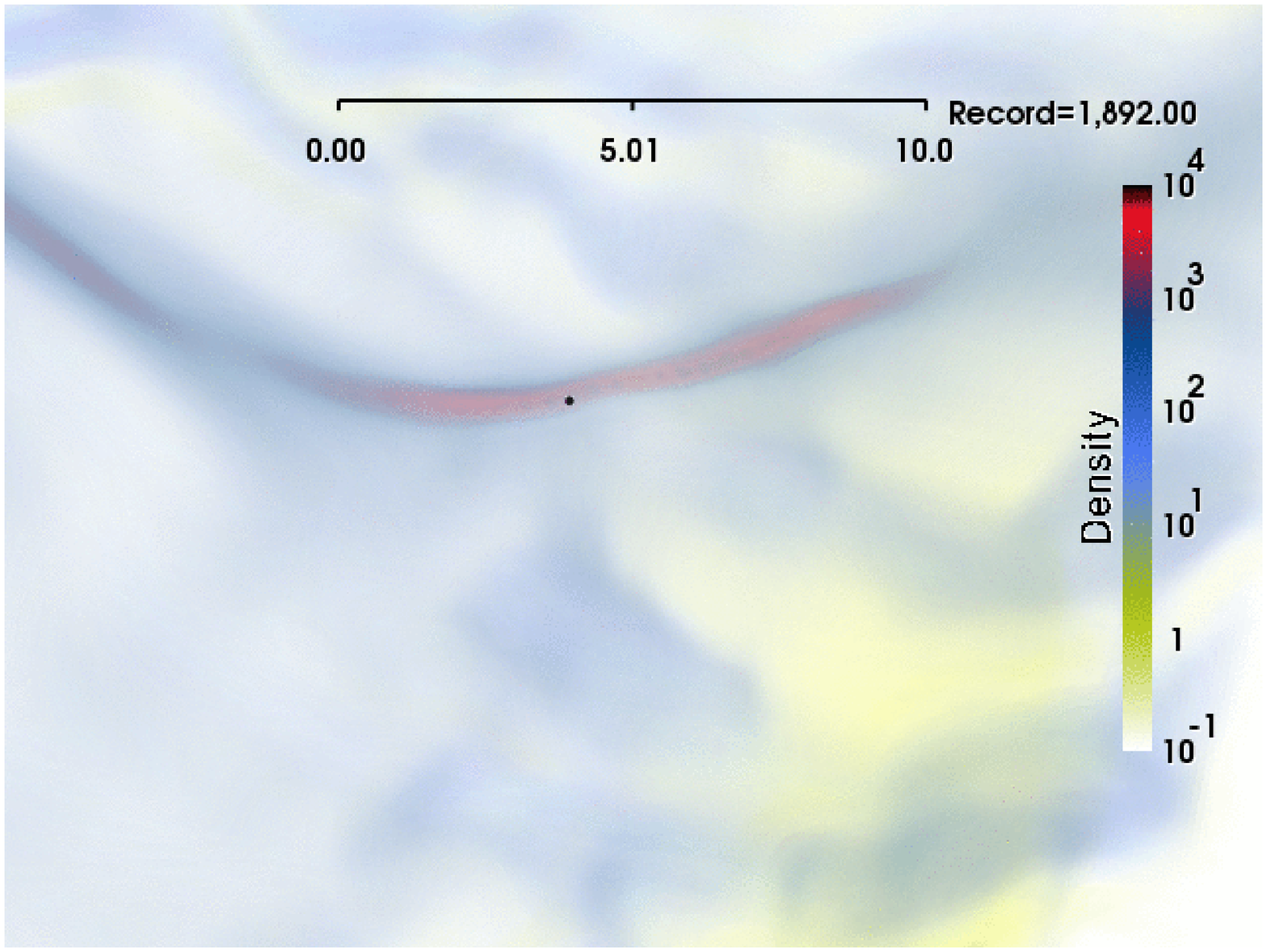}
 \includegraphics[width=0.49\textwidth]{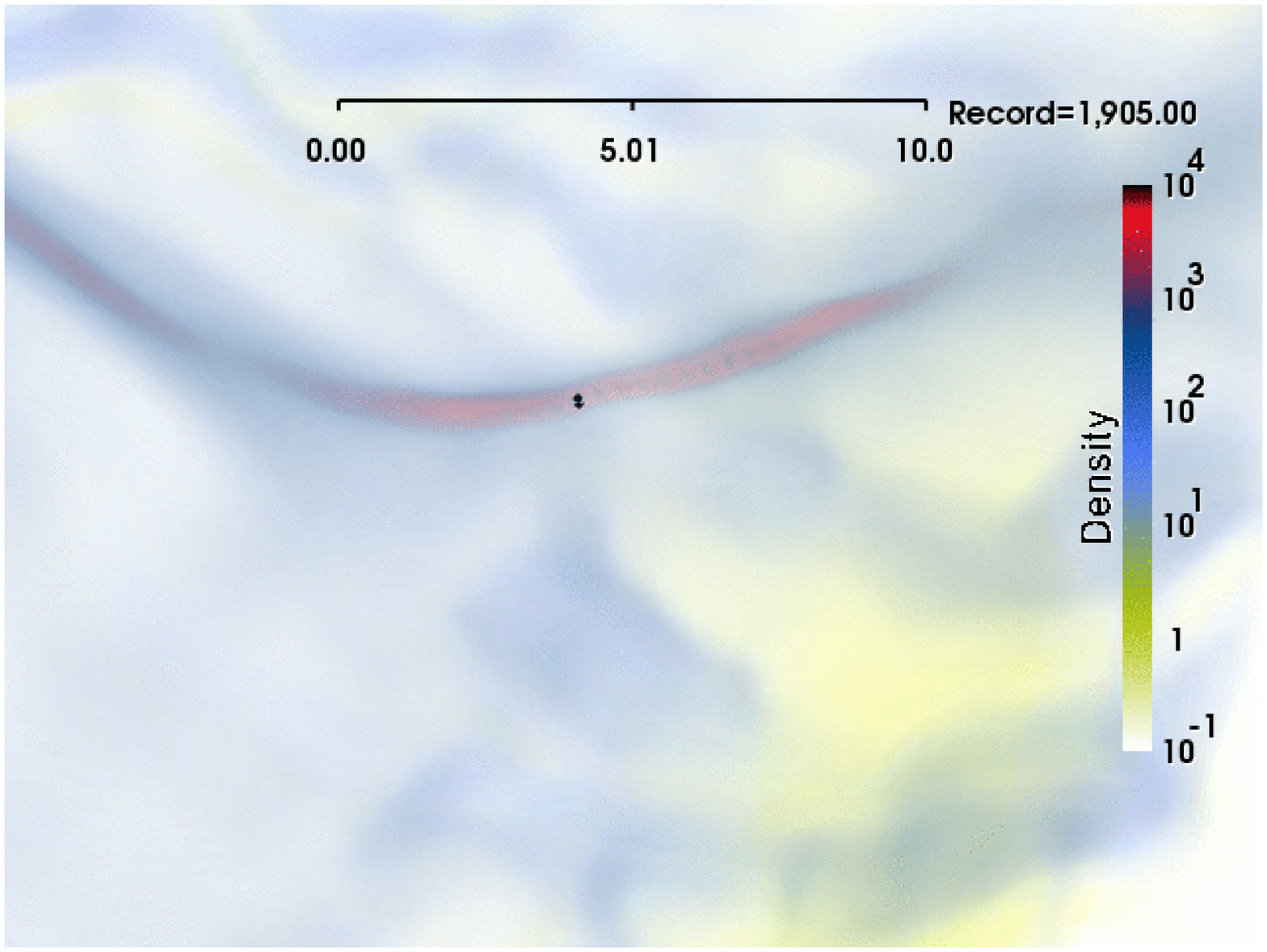}
 \includegraphics[width=0.49\textwidth]{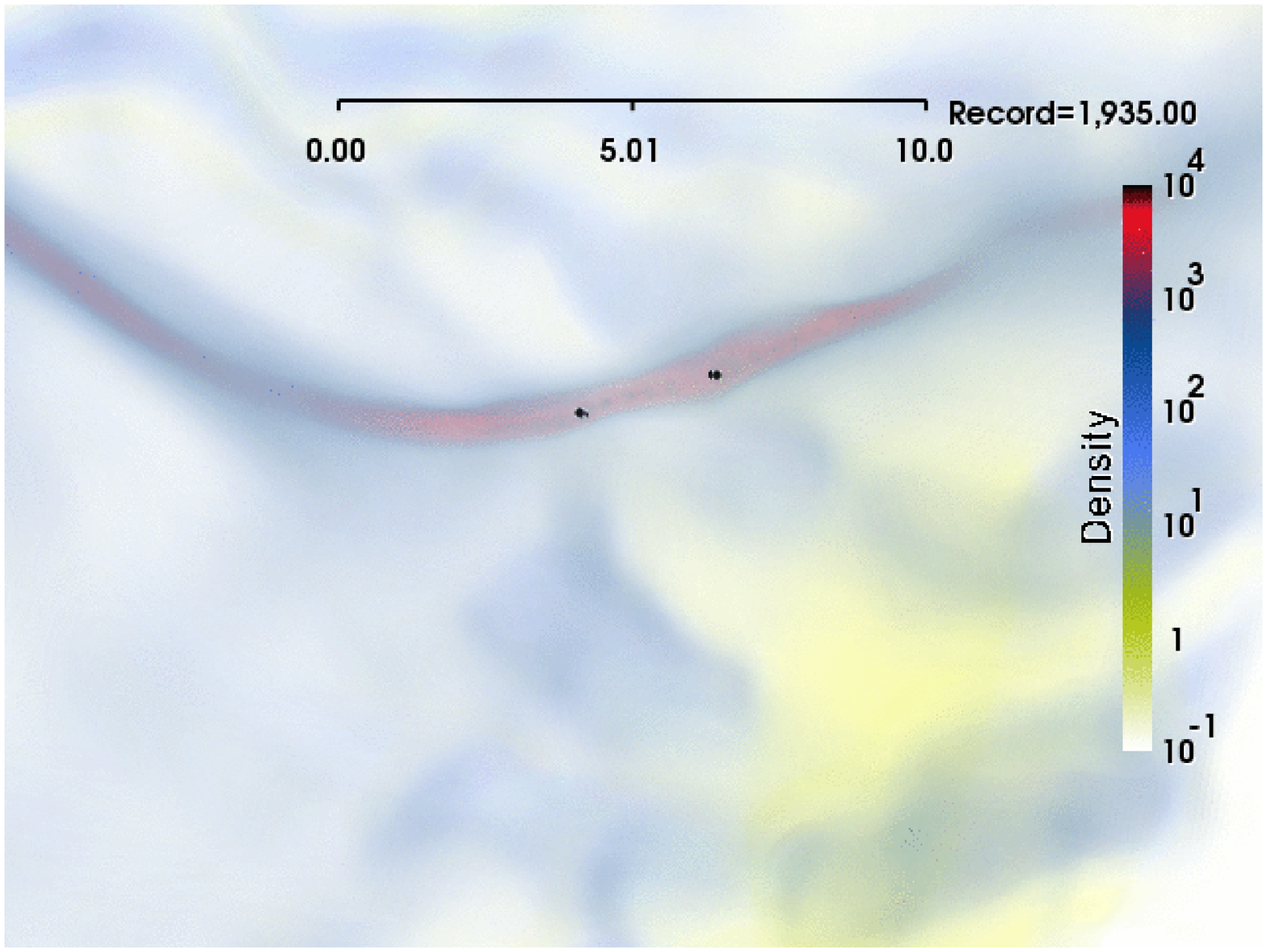}
 \includegraphics[width=0.49\textwidth]{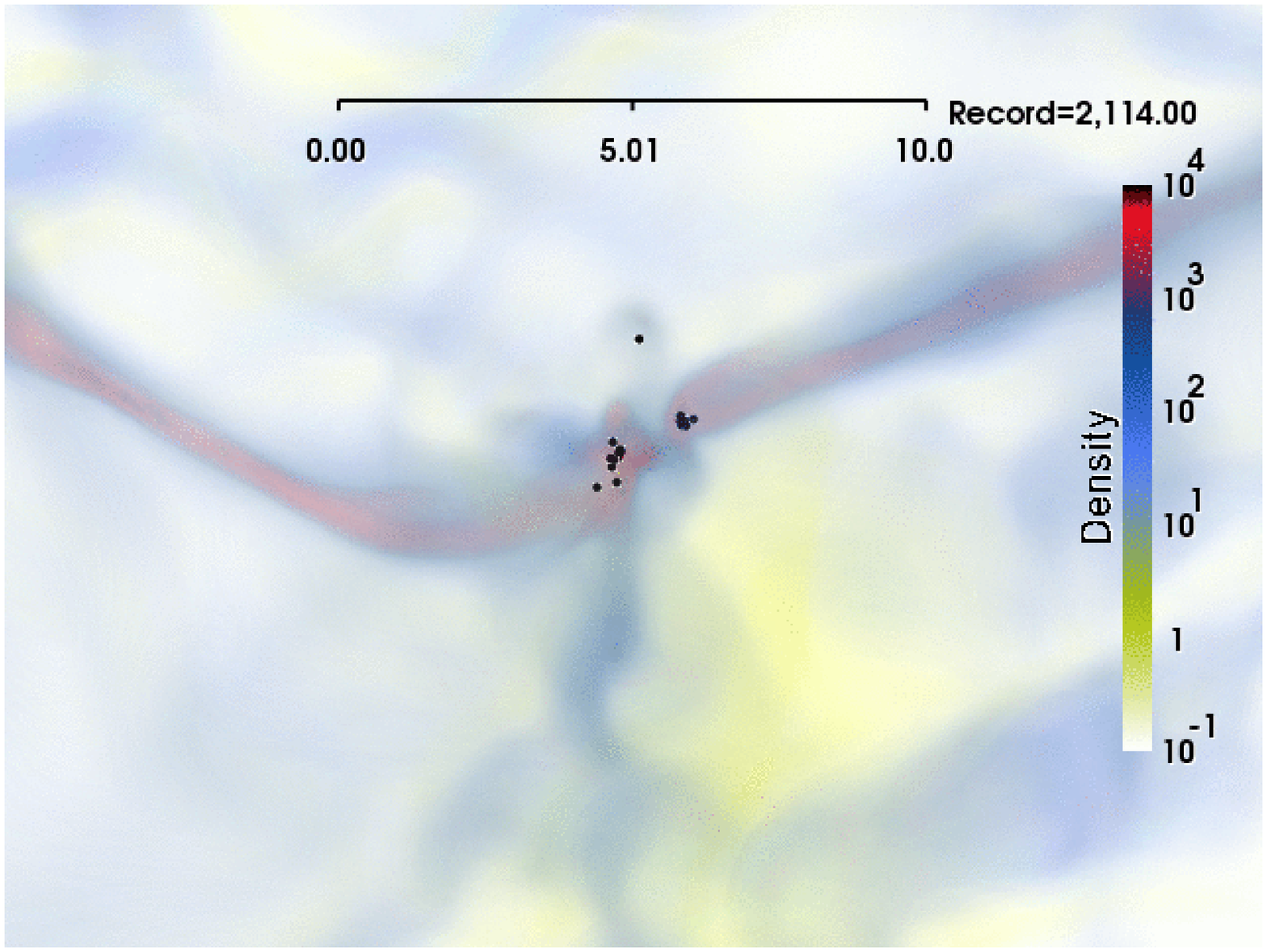}
 \includegraphics[width=0.49\textwidth]{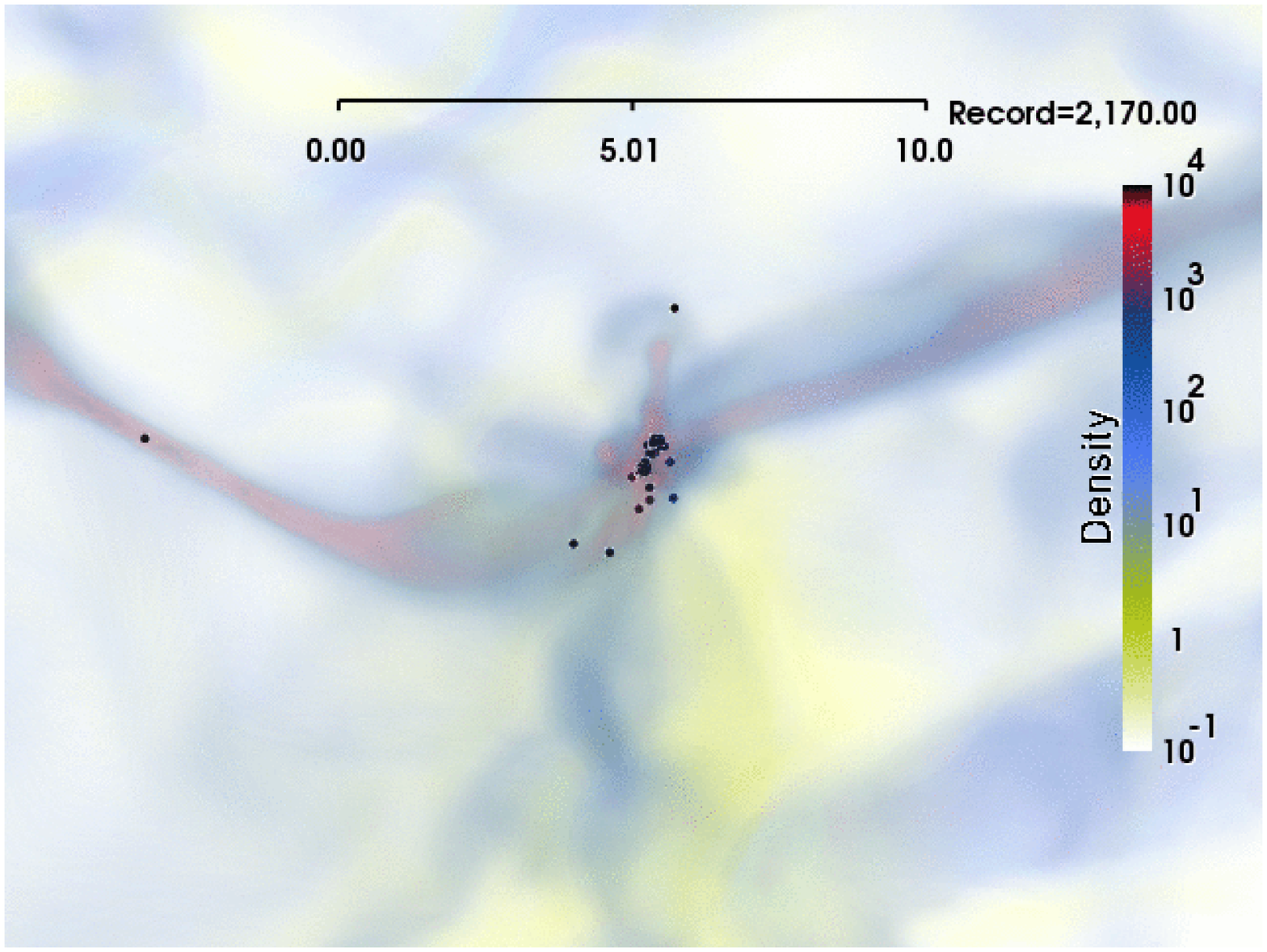}
 \includegraphics[width=0.49\textwidth]{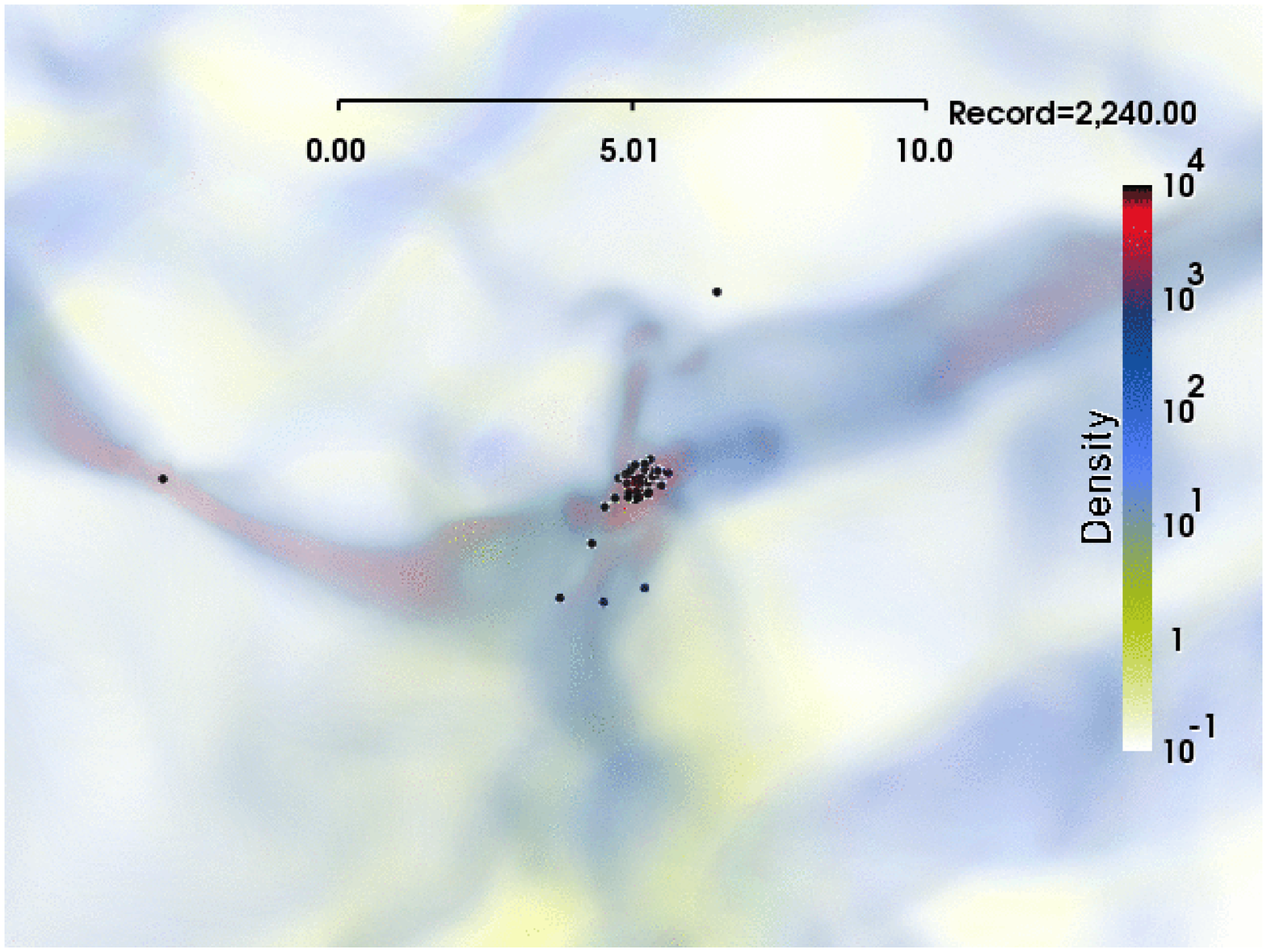} \caption{Six
 projected images zooming in around the intermediate-mass cluster at
 $t=$ 18.92, 19.05, 19.35, 21.14, 21.70, and 22.40 Myr, showing the
 hierarchical assembly of the cluster by the merging of groups that form
 along a filament, which feeds the main clump. The ruler shows a scale
 of 10 pc, and the ``Record'' numbers indicate time in units of $10^4$
 yr. The color scheme denotes the {\it volume} density of the gas
 represented in the image, with higher transparency given to
 lower-density gas, so that the highest-density gas appears purple,
 because it is shown in red, surrounded by a layer of blue-colored
 material.  See text for a description of the events.}
\label{fig:imgs_clus2}
\end{center}
\end{figure*}

The crucial effect of the GHC scenario is that the filaments constitute
part of the large-scale gravitational collapse, funneling gas into the
cores within them, as observed by \cite{GV14} in a numerical simulation
of cloud formation and collapse. Moreover, these authors observed a
hierarchy of collapses within the filaments, so that {\it small clumps,
which are collapsing locally, and sometimes forming stars already, are
themselves falling onto larger-scale ones}, similarly to the ``conveyor
belt'' scenario proposed by \citet{Longmore+14} for the gas stream in
the Central Molecular Zone.
This is illustrated in Fig.\ \ref{fig:imgs_clus2}, which shows a series
of snapshots around the second most massive cluster (hereafter,
``Cluster 2'', located in the upper left part of the cloud) among those
seen in Fig.\ \ref{fig:sim_imgs}, from $t=18.92$ Myr to $t=22.40$ Myr.
From $t=$ 18.92 to 19.05 Myr it can be seen that new stars have formed
in the first group. At $t=19.35$ Myr, a second group is seen to have
formed at a distance $\sim 2.5$ pc from the first. By $t=21.14$ Myr,
this second group is seen to have approached the first, being at a
distance $\sim 1.5$ pc from it, and to have merged with it by $t =
21.70$ Myr. Also, at this time, a third group is seen to begin forming
in the far left part of the filament, containing only one star at that
time, although by $t=22.83$ Myr it already contains several stars (not
shown). 

The effect of the feedback is also worth noting. Already by $t=21.14$
Myr, the section of the filament connecting the two clusters is seen to
be partially disrupted, and some gas is being expelled from the clump
containing the main group. However, accretion along the filament is seen
to have replenished this section of the filament at $t= 21.70$ Myr,
although the other side of the filament is now seen to be in the process
of dispersal. Finally, by $t=22.40$ Myr, the filament is seen to be in
the process of dispersal on both sides of the cluster.

\begin{figure*}
\begin{center}
 \includegraphics[width=0.49\textwidth]{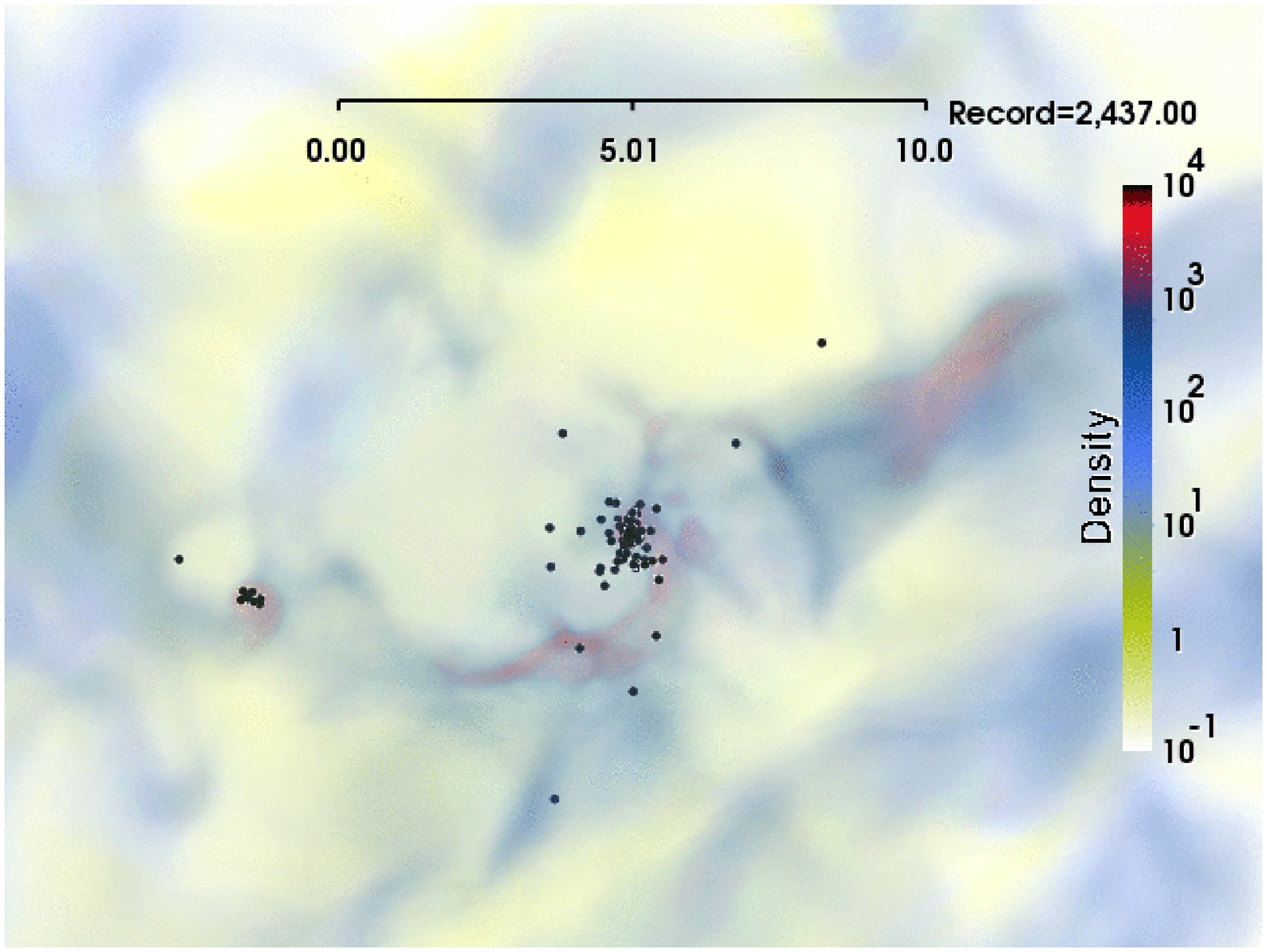} 
 \includegraphics[width=0.49\textwidth]{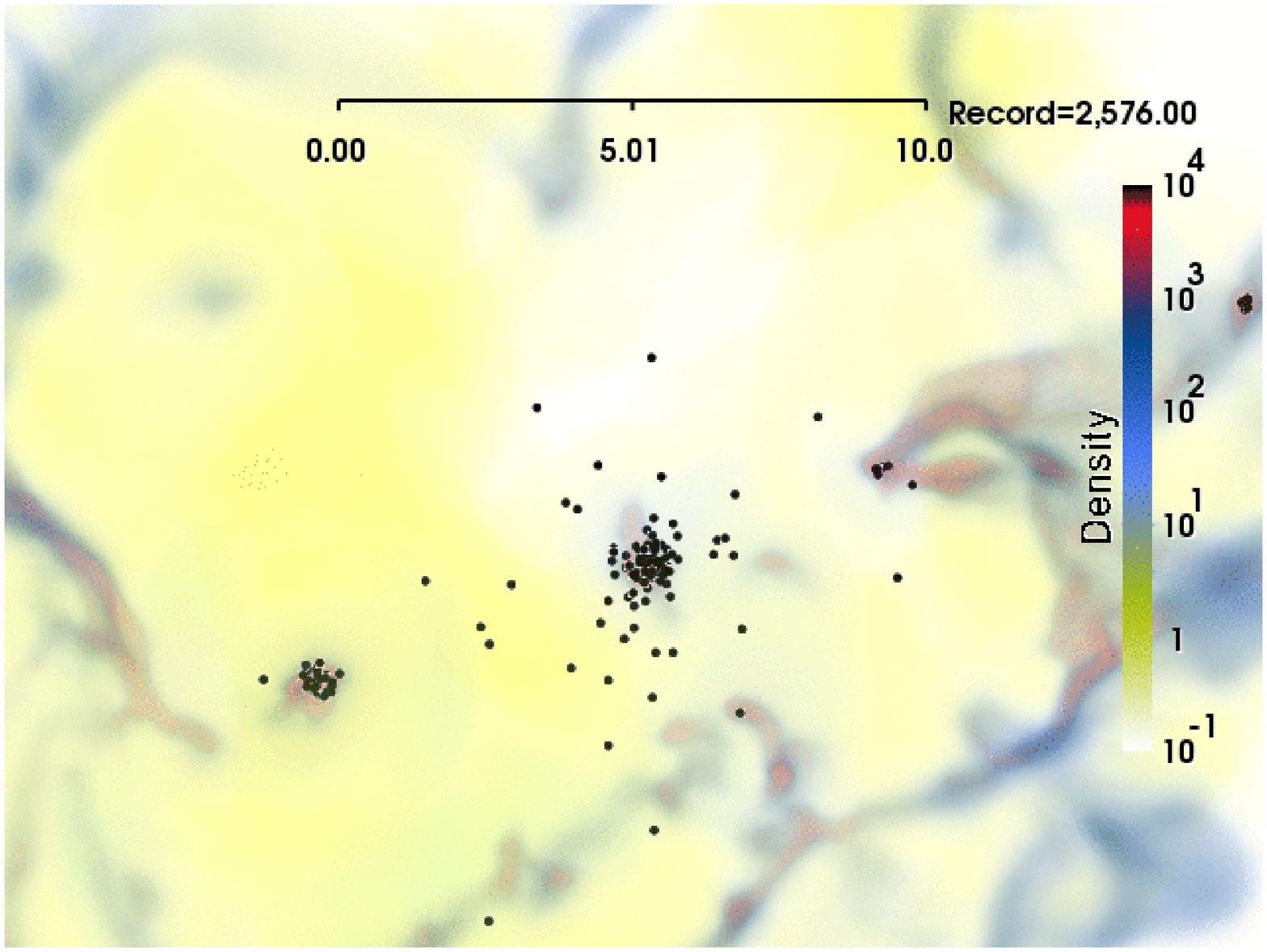} 
 \includegraphics[width=0.49\textwidth]{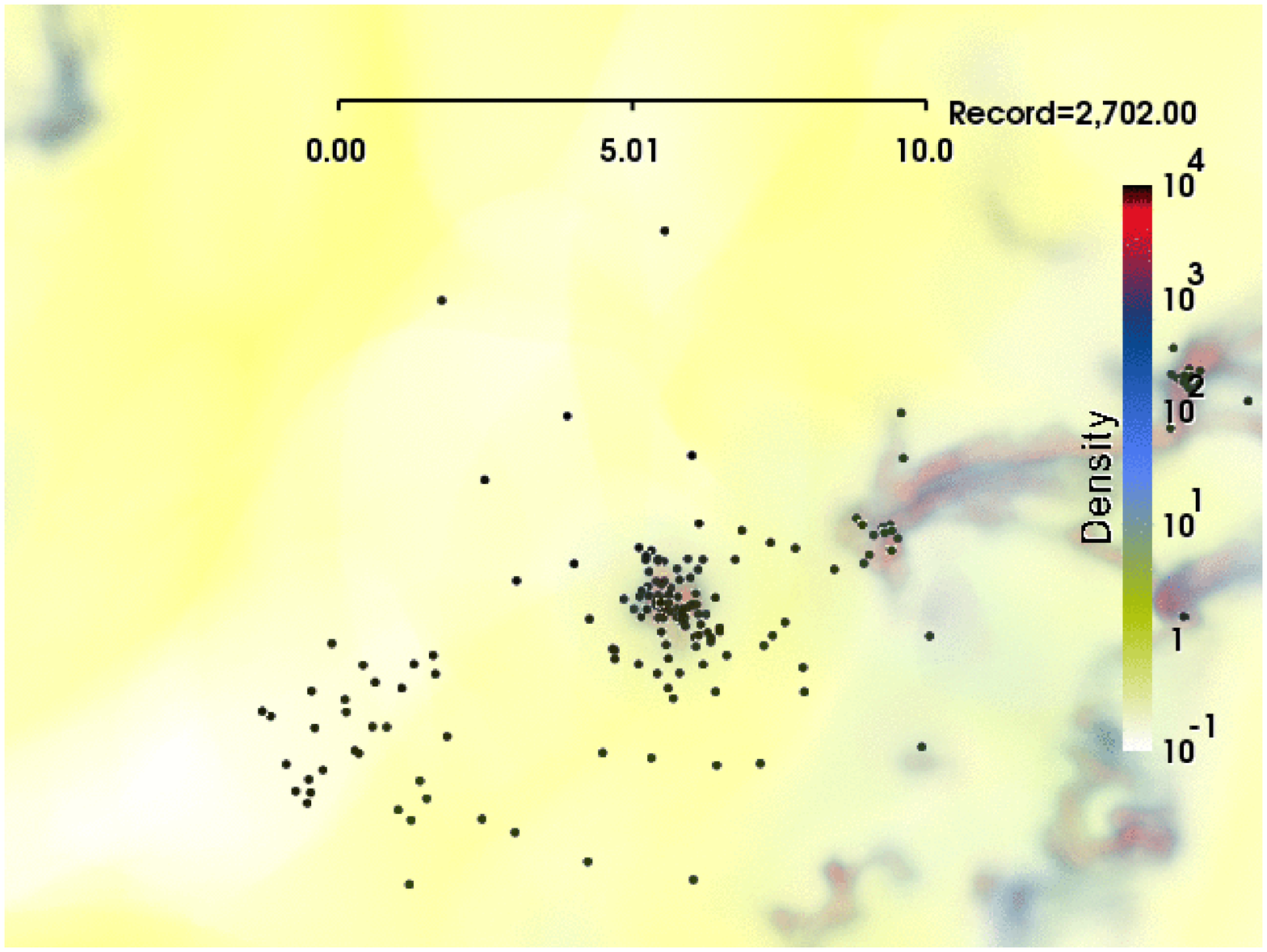} 
 \includegraphics[width=0.49\textwidth]{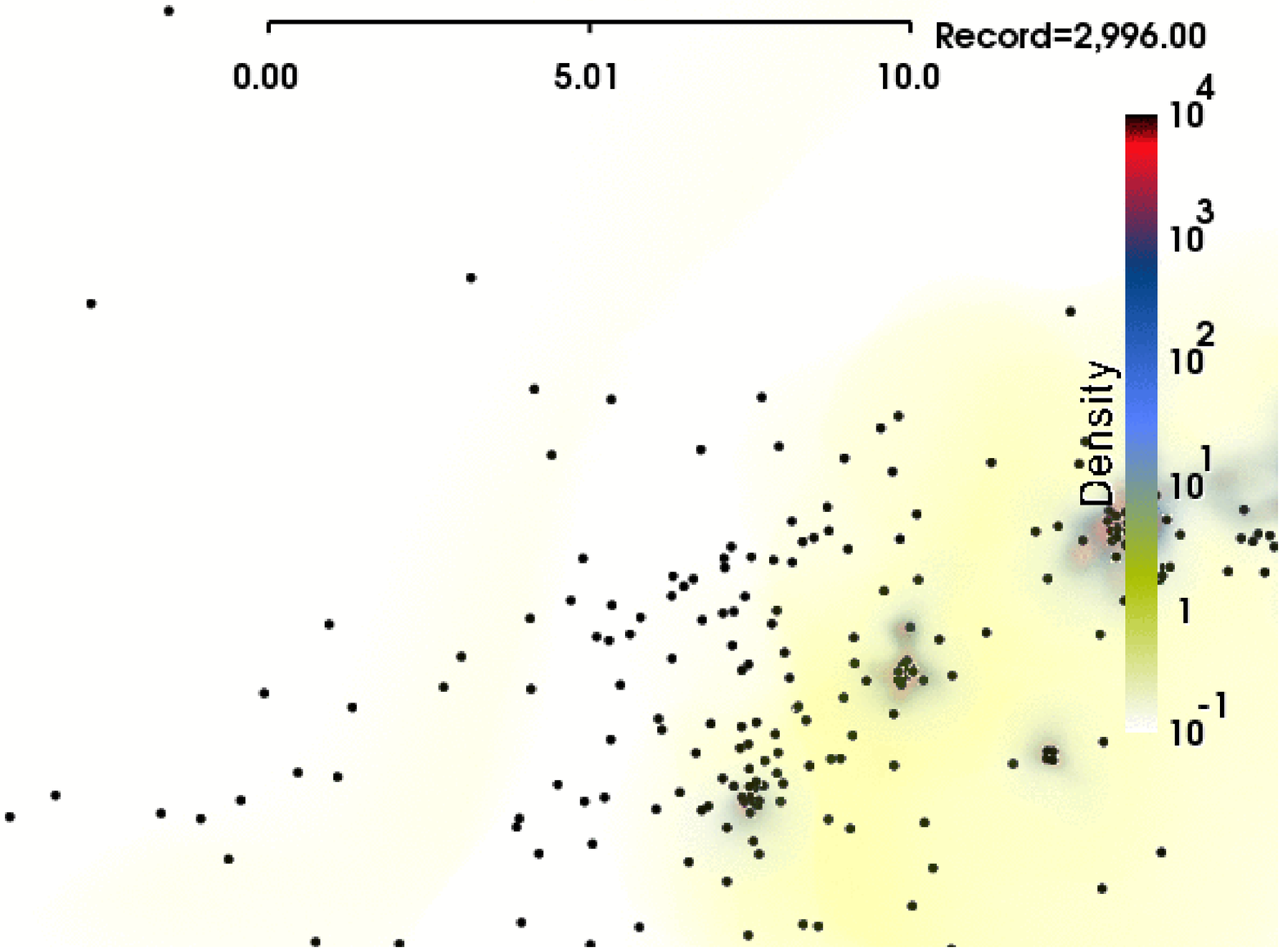} 
 \caption{Four projected images zooming in around the intermediate-mass
   cluster at $t=$ 24.37, 25.62, 27.02, and 29.96 Myr, showing the
   destruction of filament, formation of an {\sc Hii} region, and
   finally the dispersal of the gas. See text for a description
   of the events. Labels are as in Fig.\ \ref{fig:imgs_clus2}.}
\label{fig:imgs_clus2b}
\end{center}
\end{figure*}

Figure \ref{fig:imgs_clus2b} then shows the late stages of the evolution
of this cluster. By $t=24.37$ Myr the filament is seen to have been
dispersed out to distances $\sim 10$ pc, and by $t=25.62$ Myr, a large
\hii\ region, of diameter $\sim 15$ pc has formed, although the group on
the left retains some of its gas, and a new group has formed in an
``pillar'' that constitutes the remainder of the filament. Also, along
this filamentary structure, yet another group is seen to have formed (at
the far right of the image). By $t=27.02$ Myr, most of the dense gas has
been cleared from the region, and the cluster is seen to consist of
four groups, two of them almost devoid of gas (``naked'') and two of
them still embedded in their respective clumps.

 From the above description, it becomes clear that the clusters in this
simulation are assembled by means of a hierarchical process, in which
subunits formed at slightly different locations and times merge as they
fall into a large-scale potential well, {\it feeding the central clump
with both stars and gas}. In the following sections we discuss various
implications of this mechanism.

\subsection{Subgroup evolution and merger} 
\label{sec:exp}

\begin{figure*}
 \includegraphics[width=0.49\textwidth]{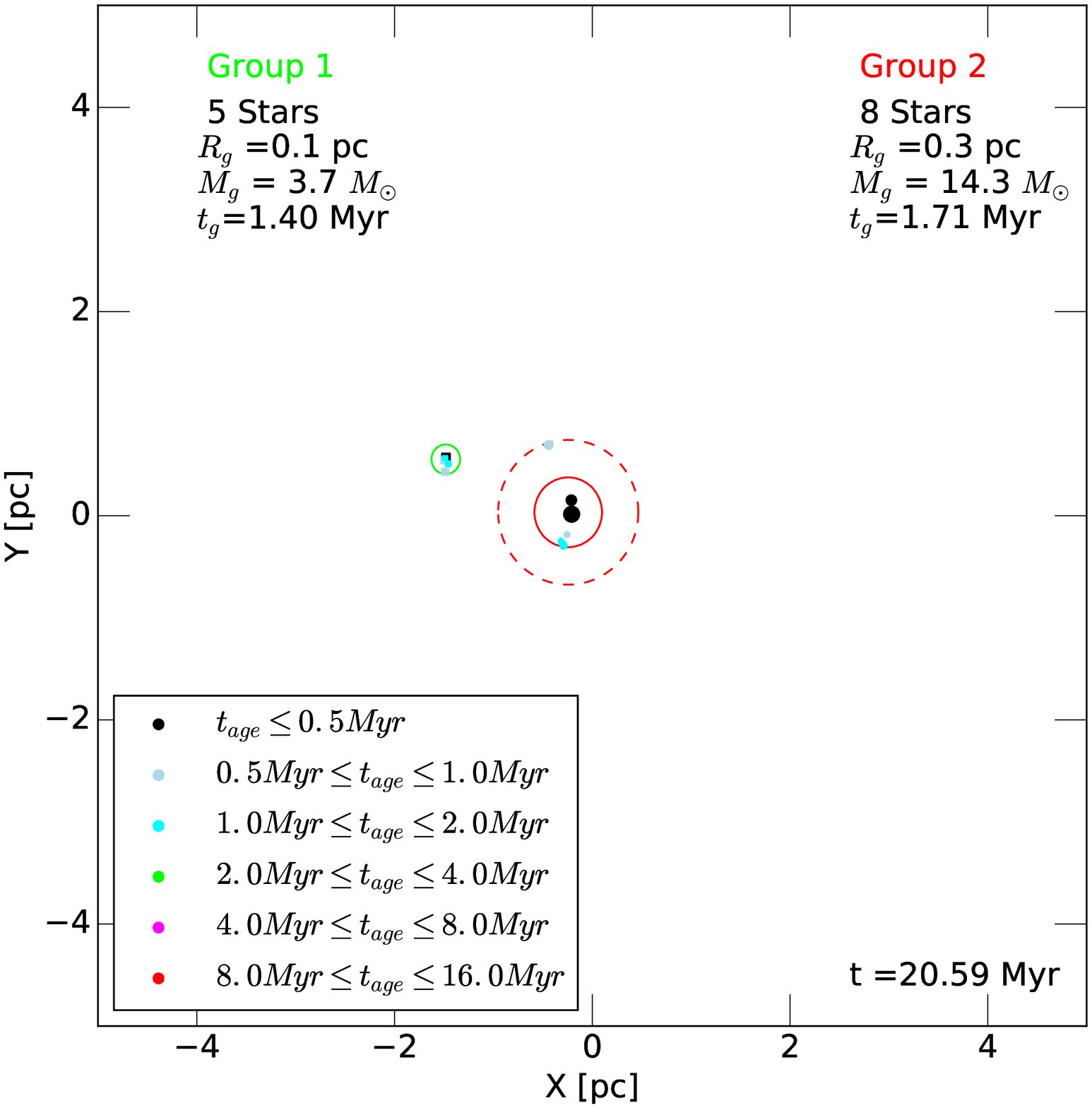}
 \includegraphics[width=0.49\textwidth]{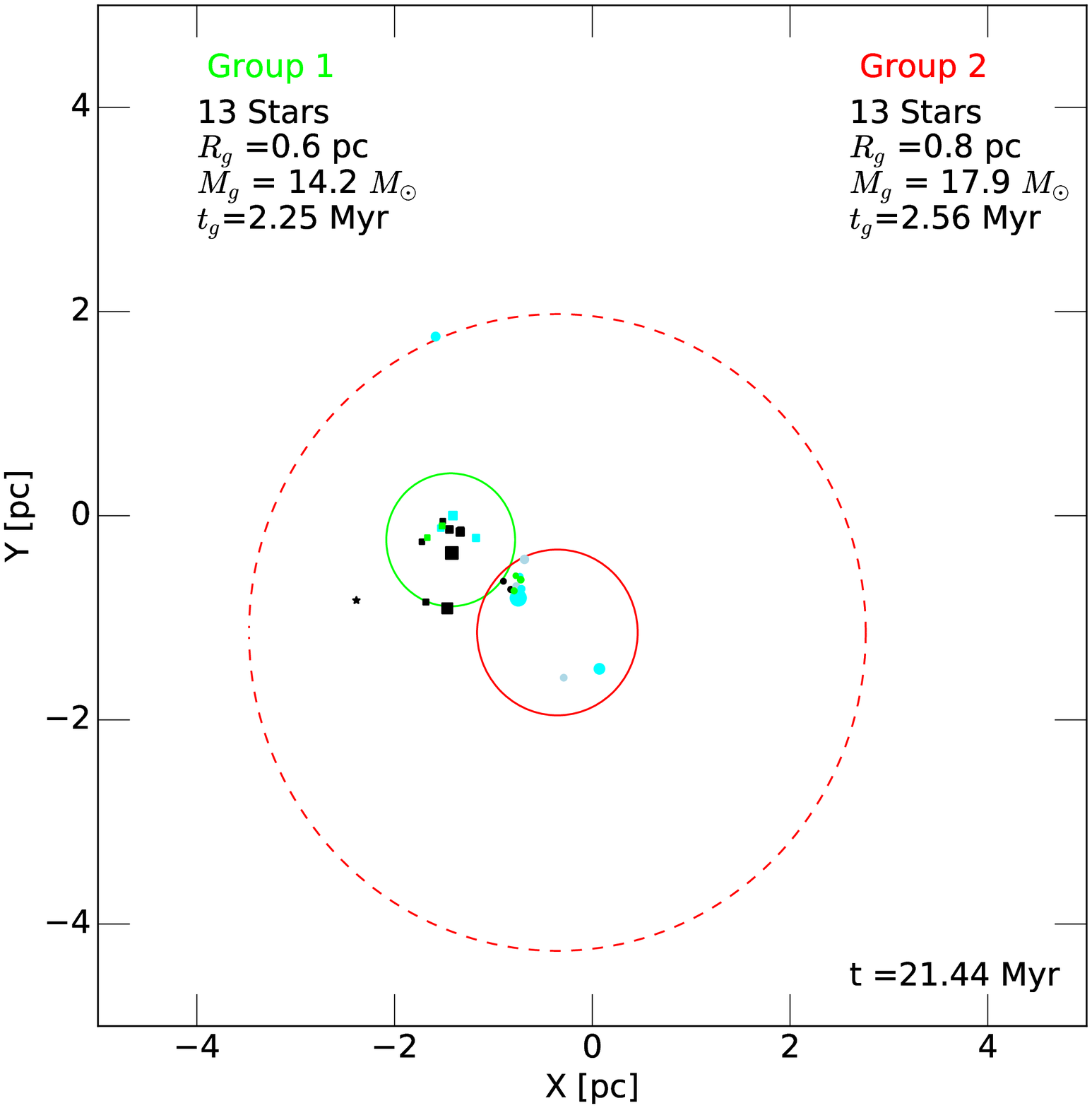}
 \includegraphics[width=0.49\textwidth]{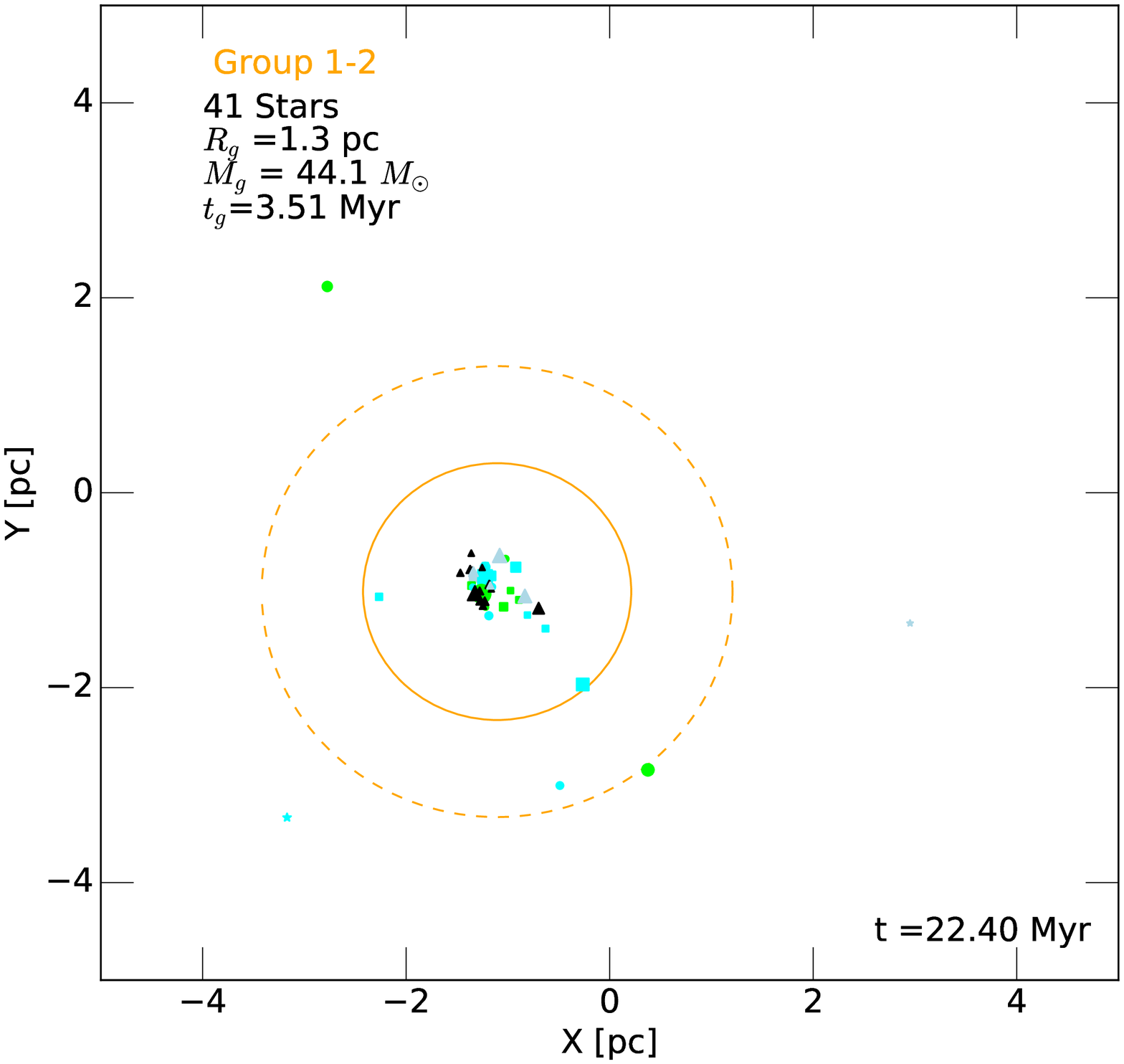}
 \includegraphics[width=0.49\textwidth]{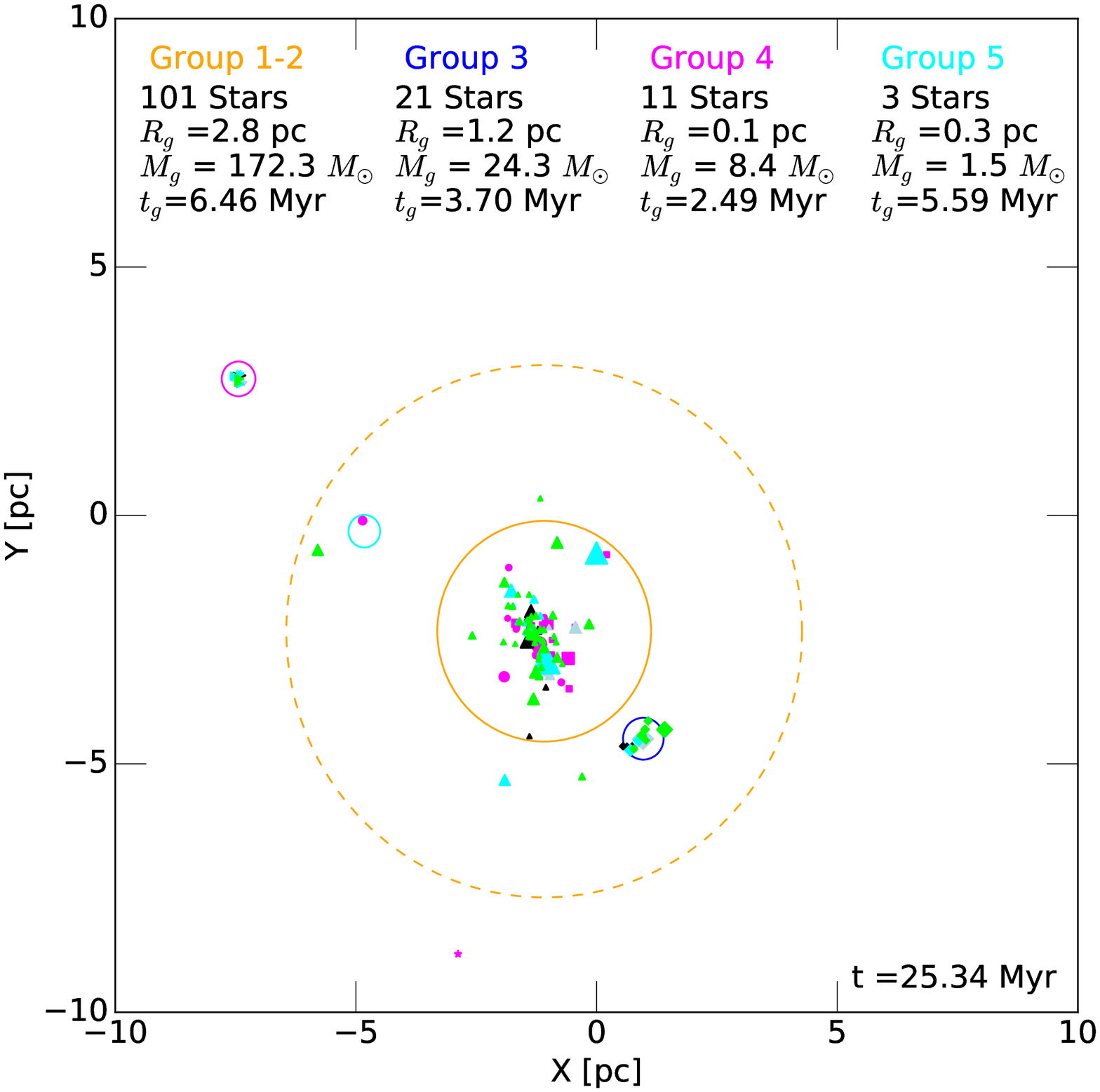}
 \caption{Four snapshots of ``cluster 2'' at times $t=20.59$, 21.44,
   22.44 and 25.34 Myr in the simulation (respectively, {\it top left,
     top right, bottom left}, and {\it bottom right} frames), showing
   the merging of two sub-groups to form a larger one, while
   simultaneously they expand and stars continue to form in their parent
   clumps. Group membership is defined as described in Sec.\
   \ref{sec:cluster}. Square symbols represent stars that belong to
   Group 1, and filled circles represent stars belonging to Group 2. The
   two groups merge at $t=21.56$. Triangles represent stars born in
   the new Group 1-2. By $t=25.34$ Myr, three more groups, named Groups
   3, 4, and 5, are seen to have formed, lying approximately in a
   straight line, because they formed in the filament feeding the main
   clump. Colors represent stellar ages as indicated in the box in the
   {\it top left} frame. Solid circles represent the group's size out to
   the most distant star, and dashed circles represent the size out to
   the second most distant star.  Finally, the symbol sizes
   qualitatively represent the stellar particles' masses.}
\label{fig:cluster_evol}
\end{figure*}

The resulting evolution of Cluster 2 is illustrated in Fig.\
\ref{fig:cluster_evol}, which shows the projected positions of the
stellar particles on the $(x,y)$ plane at times $t=20.59$, 21.44,
22.40, and 25.34 Myr in the {\it top left, top right, bottom left, and
  bottom right panels}, respectively.

In this cluster, stars begin to form at $t
\approx 19.2$ Myr and Fig.\ \ref{fig:cluster_evol} shows that they are
assembled into a cluster in such a way that at 
$t=20.59$ Myr we identify two subgroups (hereafter, Groups 1 and 2; see
the {\it top left panel} of Fig.\ \ref{fig:cluster_evol}) of what will
later become a larger, merged group (Group 1-2). We use square symbols to
identify stars that belong to Group 1, and filled circles for stars
belonging to Group 2. Furthermore, we color them according to their age
as shown in the labels and qualitatively represent their masses by the
symbol size, with larger symbols indicating larger masses. The green and
red circles represent the group radii at each time, computed as
explained in Sec.\ \ref{sec:cluster}, with the dashed circles
representing the size out to the most distant star, and the solid
circles representing the distance to the second most distant star. In
what follows, we use the distance to the second most distant star as the
radius of Group 1-2, because the first most distant one is clearly ``running
away'' from the group. Finally, we use larger symbols to represent more
massive stars, and smaller symbols to represent lower-mass stars. Thus,
even at these early stages, we can see that {\it the more massive and
  younger stars tend to be near the centers of the groups}.

From Fig.\ \ref{fig:cluster_evol} it is clearly seen that, at these
early stages, the groups are undergoing expansion, and that star
formation continues within them.  By $t=21.44$ Myr, both groups have
similar sizes, masses, and number of stars, and they have undergone a
large enough expansion that their merger appears inevitable. The
formerly ``most distant'' star of Group 2 is now far away from the
center of either group. By our definition, the merger occurs at $t =
21.56$ Myr, forming the new Group 1-2. We have kept the
circular and square symbols to indicate the original group membership of
the stars in the merged group. Triangles now represent stars born in the
new Group 1-2. Thus, this new group contains stars that were born at
somewhat different times and locations.

For later times, we continue to track the evolution of this and other,
new groups appearing in a 10-pc box around Group 1-2, and by $t=25.34$
Myr, we now find a system of groups with different ages, masses, and
sizes ({\it bottom right panel} of Fig.\ \ref{fig:cluster_evol}). These
groups lie approximately in a straight line, because they formed in the
filament feeding the main clump. Eventually, as these groups expand and
approach each other, they finally become part of what we refer to as the
whole Cluster 2, as seen in the {\it bottom right panel} of Fig.\
\ref{fig:imgs_clus2b}, although the subunits are still partially
distinguishable at that time ($t=29.96$ Myr).

\subsection{Star formation rate and delayed massive-star formation}
\label{sec:SFR} 

\begin{figure}
 \includegraphics[width=0.49\textwidth]{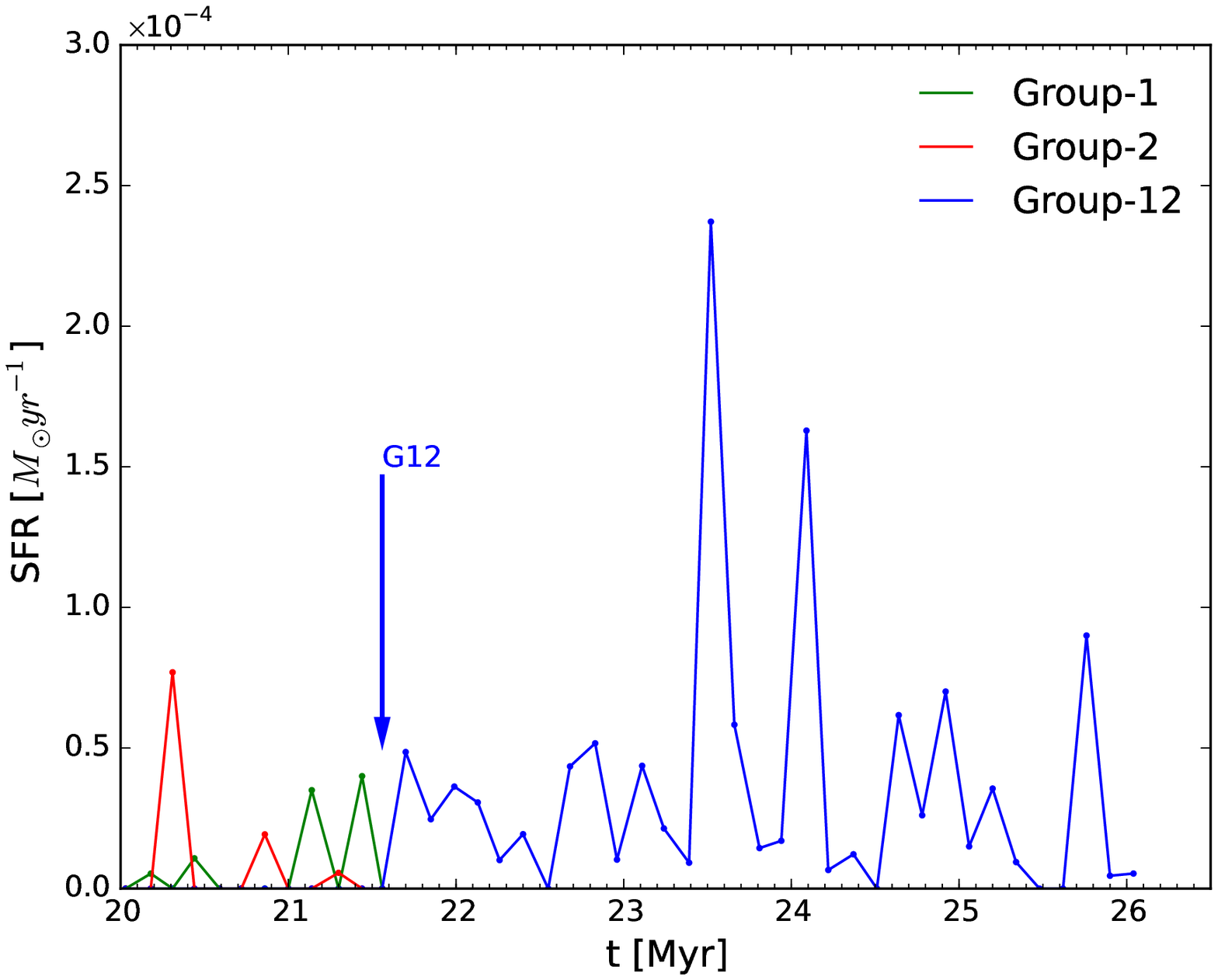} 
\caption{Evolution of the SFR for Groups 1, 2 and 1-2.}
\label{fig:sfr_evol}
\end{figure}

As mentioned in the Introduction, one fundamental implication of the GHC
scenario is that the SFR of star-forming clouds must increase in time
until the time when massive stars begin to appear and to destroy their
parent MC \citep{ZA+12, ZV14}. Figure \ref{fig:sfr_evol} shows the
evolution of the SFR in Groups 1 and 2, as well as in the merged Group
1-2. Except for Group 2, in whose parent cloud the SFR starts
relatively high and then decreases, for Groups 1 and 1-2 the trend of
increasing SFR is observed. Moreover, the SFR for the merged Group 1-2
is much larger than the maximum value reached in Group 2 before the
merger, so the increase of the SFR also holds for the system of Groups 1
and 2 combined, before and after their merger.

\begin{figure*}
 \includegraphics[width=0.49\textwidth]{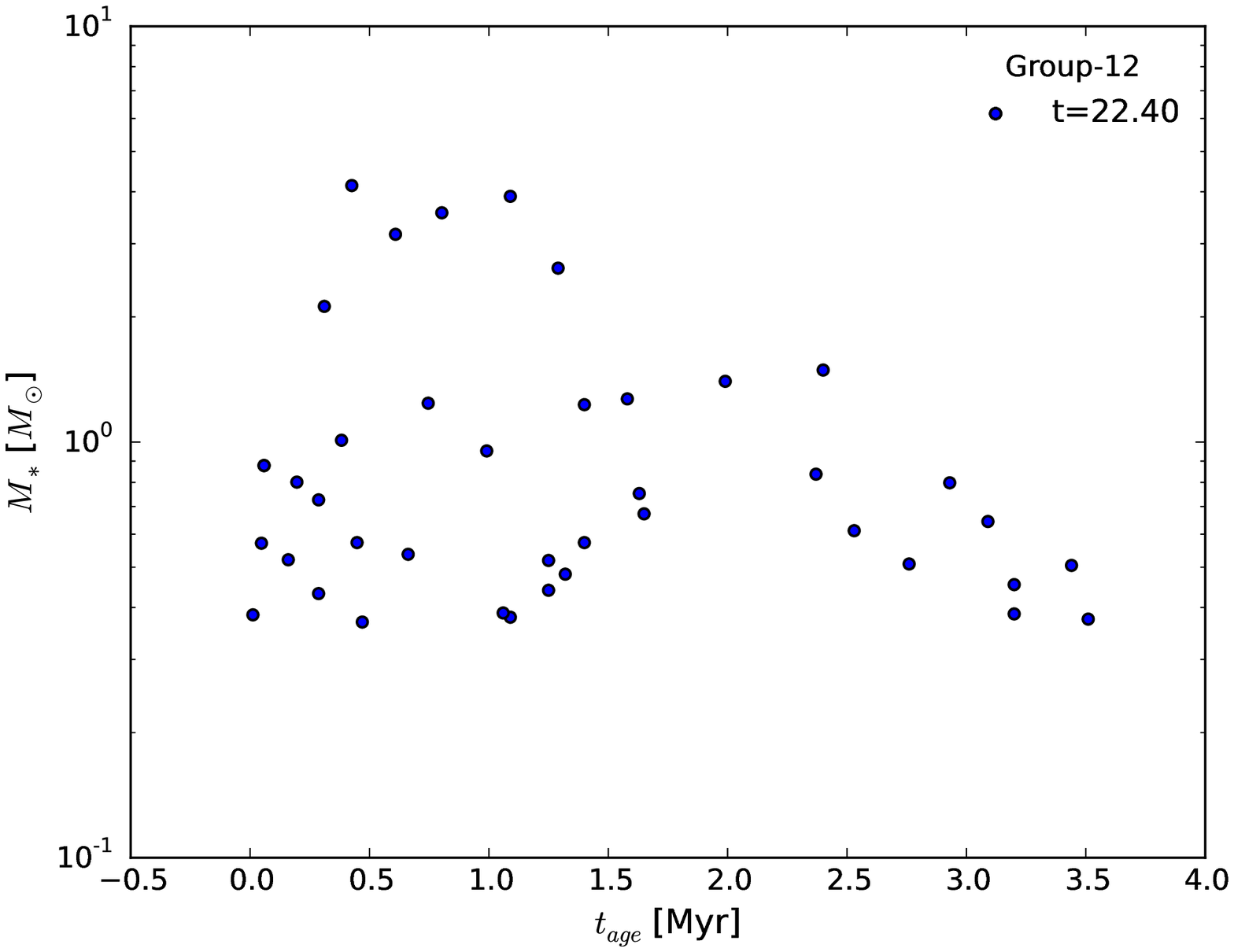}
 \includegraphics[width=0.49\textwidth]{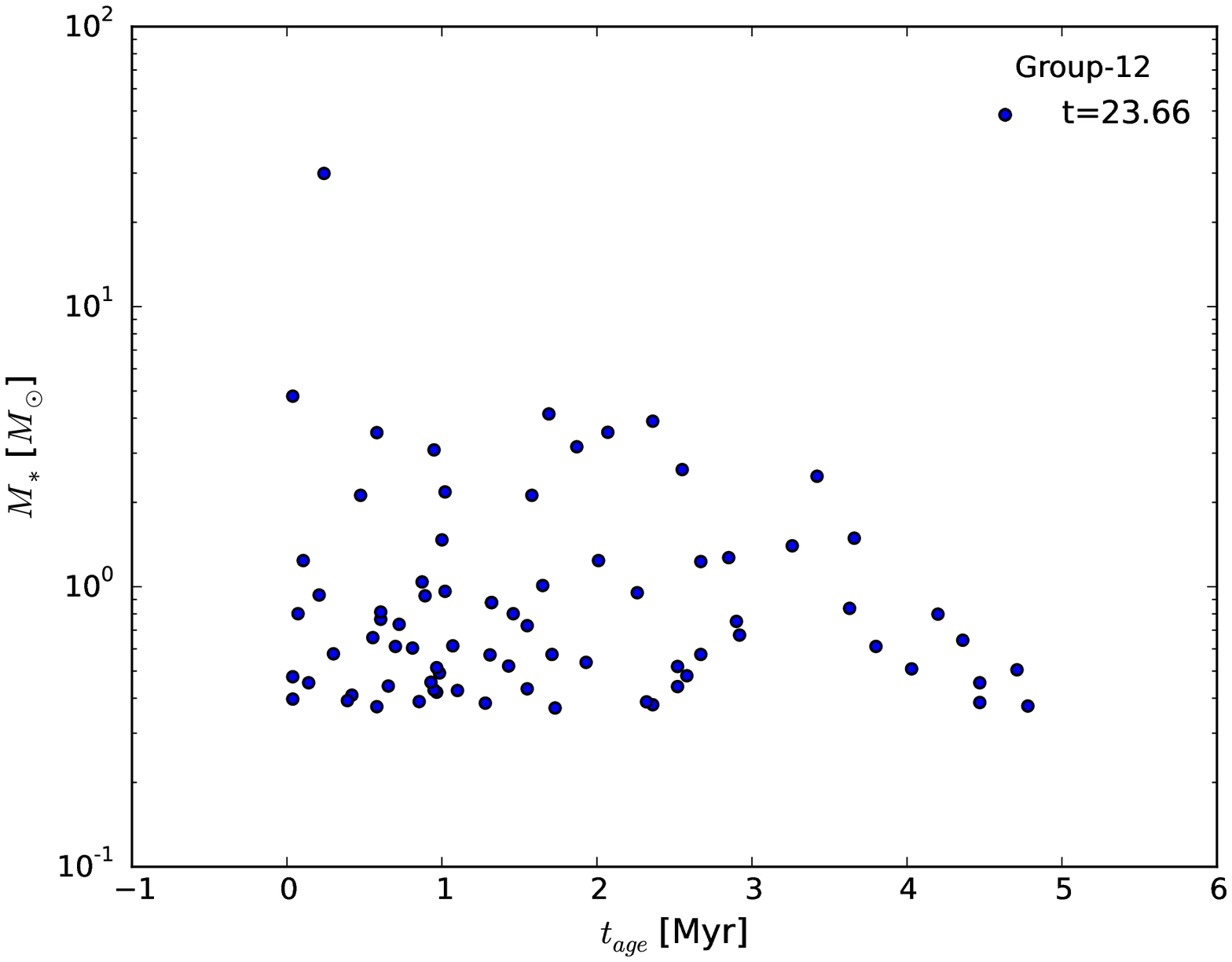}
 \caption{Mass {\it vs.} age of the cluster stars at times $t=22.40$
   ({\it left}) and 23.66 Myr {\it right}). The oldest stars are clearly
   seen to have masses near the minimum value. The younger stars,
   however, can be either young or old.  Recall that the minimum mass of
   $0.39 \Msun$ is artificially imposed by our stellar particle
   formation scheme.}
 \label{fig:age-mass}
\end{figure*}

Assuming that massive stars do not form in a region until the SFR is
locally high enough to sample the IMF up to large masses, the increase of
the SFR implies that the massive stars should tend to appear late in the
evolution of star-forming regions. This is illustrated in Fig.\
\ref{fig:age-mass}, which shows the mass of the stars {\it vs.} their
age at $t=22.40$ and $t=23.66$ Myr. At both times, it is clearly seen
that the oldest stars have the lowest masses, and that, as the age
decreases, the range of stellar masses extends to larger values.

\begin{figure}
 \includegraphics[width=0.49\textwidth]{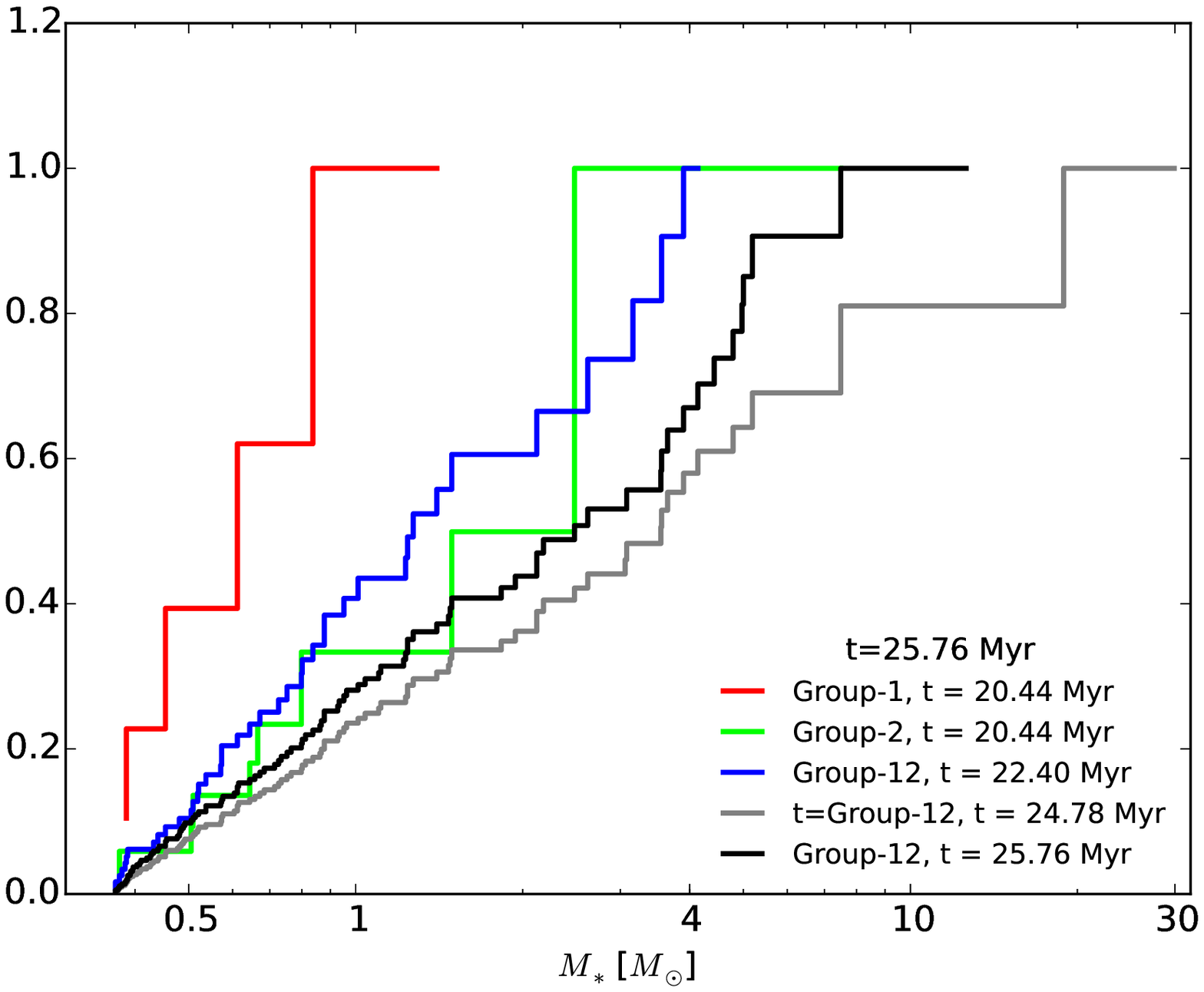}
 \caption{Normalized cumulative stellar mass histogram for Groups 1, 2,
   and 1-2 at various times. As time proceeds, a larger fraction of the
   stars are seen to be massive.}
\label{fig:cum_mass_hist}
\end{figure}

This is also illustrated in Fig.\ \ref{fig:cum_mass_hist}, which shows
the cumulative mass distribution of Groups 1, 2, and 1-2 at various
times. It is seen from this figure that the relative abundance of
massive stars increases over time, especially after Groups 1 and 2 have
merged.

\subsection{Spatial and statistical distributions of age and mass}
\label{sec:age_mass_distr} 

\subsubsection{Age distributions and gas and star velocity dispersions}
\label{sec:age_distr}

The GHC mechanism implies that each star-forming site is locally undergoing
collapse while simultaneously it is falling into larger-scale potential
wells. Moreover, since the small-scale regions are forming stars on
their own, this process in turn implies that the larger scales are fed
both gas and stars (cf.\ Sec.\ \ref{sec:global_evol}) from the
surrounding infalling material. Also, the small-scale regions
fall into the larger-scale ones mostly along filaments \citep{GV14}.

An important consequence of this ``mixed'' (gas+stars) infall is that the
ensemble of stars constitutes a non-dissipative ``fluid'' that conserves
the kinetic energy of the infalling motions. Thus, the stars formed {\it
  before} reaching the large-scale collapse center should exhibit a
velocity dispersion corresponding to their motions as they were
infalling to the center of the large-scale potential well. In turn, this
implies that older stars should reach somewhat larger distances from the
center of a group or cluster that has undergone mergers than the stars
formed {\it in situ} in this merged clump, where the gas from which they
inherit their velocity dispersion has probably been shocked and thus has
lost some of the infall kinetic energy.

%

\begin{figure*}
 \includegraphics[width=0.49\textwidth]{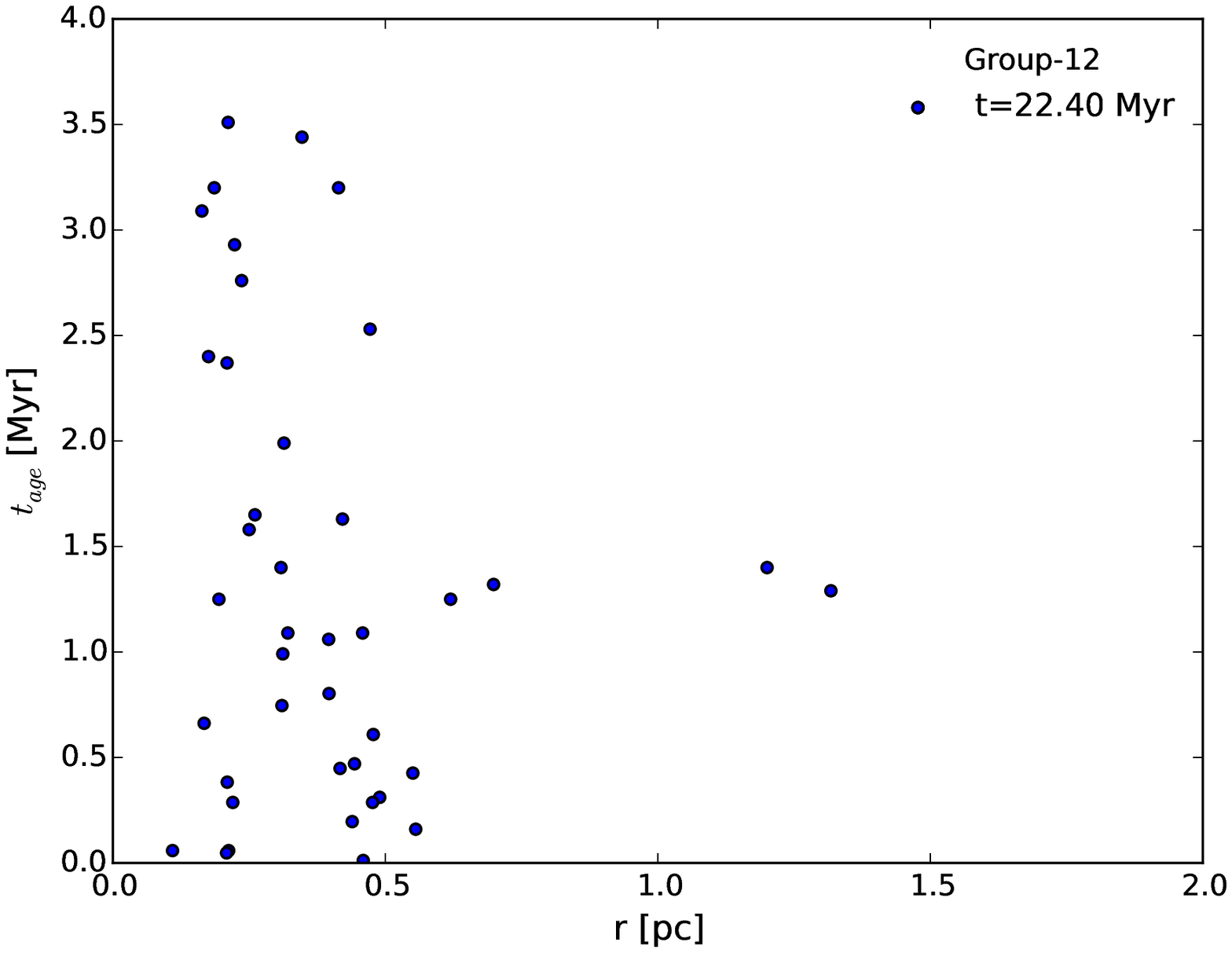}
 \includegraphics[width=0.49\textwidth]{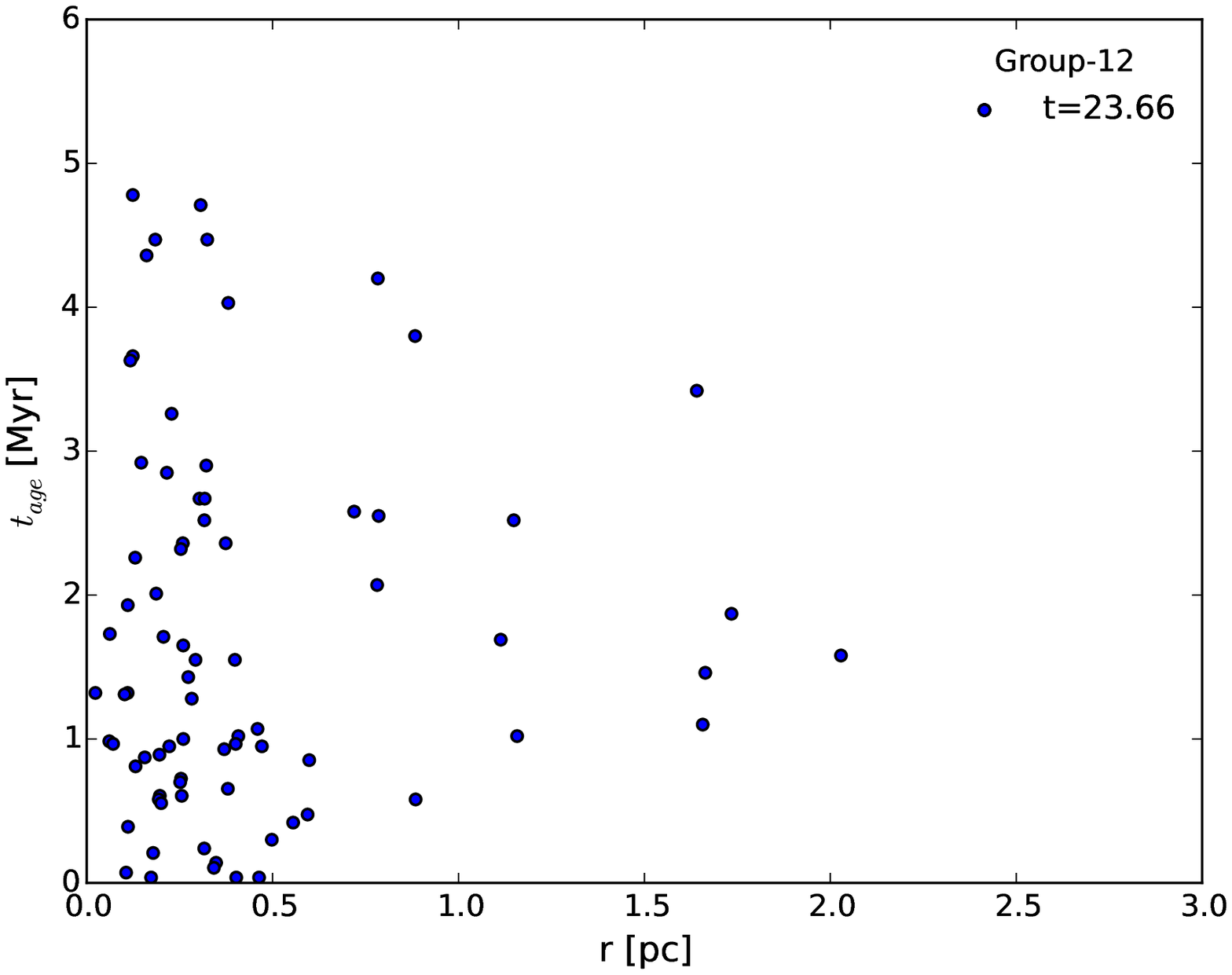}
 \caption{Radial age distribution of the stellar particles in the
 cluster at times $t=22.40$ and 23.66 Myr. At these times, the cluster
 is composed of the merger of Groups 1 and 2. The central 0.5 pc
 contains the youngest stars (from $\tage \la 0.1$ to 0.5 Myr), while no
 stars younger than $\tage \approx 1$ Myr exist at radii $r \ga 0.6$
 pc. The old ($\tage > 2$ Myr) are those that formed in Groups 1 and 2
 prior to their merger. Also, the cluster is seen to have expanded
from $t=22.40$ to 23.66 Myr from a radius $r \approx 1.4$ to 2 pc.}
\label{fig:age_profile}
\end{figure*}

Figure \ref{fig:age_profile} shows the age of the stellar particles as a
function of their distance to the center of mass of Cluster 2 at $t =
22.4$ Myr. It is seen that all of the youngest stars ($\tage \la 0.5$,
and in particular several stars with $\tage \la 0.1$ Myr), are located
within the central 0.5 pc of the cluster. This is
qualitatively consistent with the observation that the youngest stars
appear tightly clustered around core-like or filamentary star-forming
regions \citep[e.g.,] [see the discussion in Sec.\ \ref{sec:compar_obs}]
{Getman+14b, Povich+16}. 

Instead, at larger distances ($> 1$ pc), there are no particles with
$\tage \la 1.5$ Myr. These older stars are the ones that formed in
Groups 1 and 2, prior to their merging.  This can also be seen in the
{\it bottom left} panel of Fig.\ \ref{fig:cluster_evol}, which shows
that the stars farther from the center of Group 1-2 are all represented
by circles and squares (i.e., formed previously to the merger) and by
green and cyan colors, denoting ages between 1 and 4 Myr, while the
stars clustered near the center have ages $\le 0.5$ Myr, as indicated by
their black-colored symbols.

On the other hand, we also notice in Fig.\ \ref{fig:age_profile} that
the inner 0.5 pc of the cluster also contains several old stars, and in
particular, the oldest ones ($\tage \sim$ 2--3.5 Myr). This suggests
that some of the stars formed previously to the merger have already had
$N$-body interactions with the other stars in the region, and thus have
transferred their excess kinetic energy to other stars in the cluster.
This is facilitated because these old stars have low masses, so that they
are strongly affected by interactions with more massive ones. Indeed, as
discussed in Sec.\ \ref{sec:SFR} (Fig.\ \ref{fig:age-mass}), the
oldest stars have masses near the peak of the IMF, which in our case is
also the minimum stellar mass (cf.\ Sec.\ \ref{sec:SF}).

\begin{figure*}
 \includegraphics[width=0.49\textwidth]{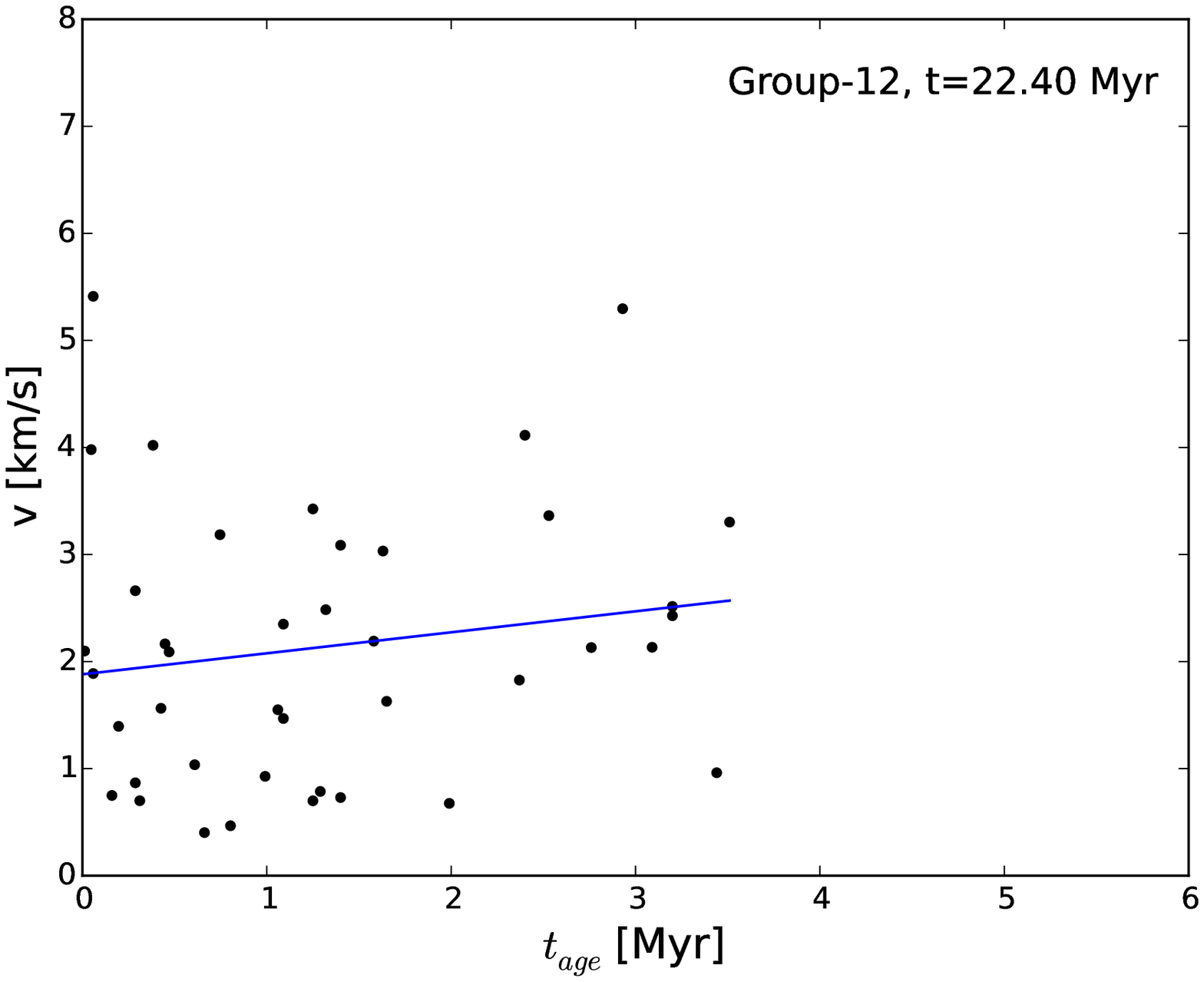}
 \includegraphics[width=0.49\textwidth]{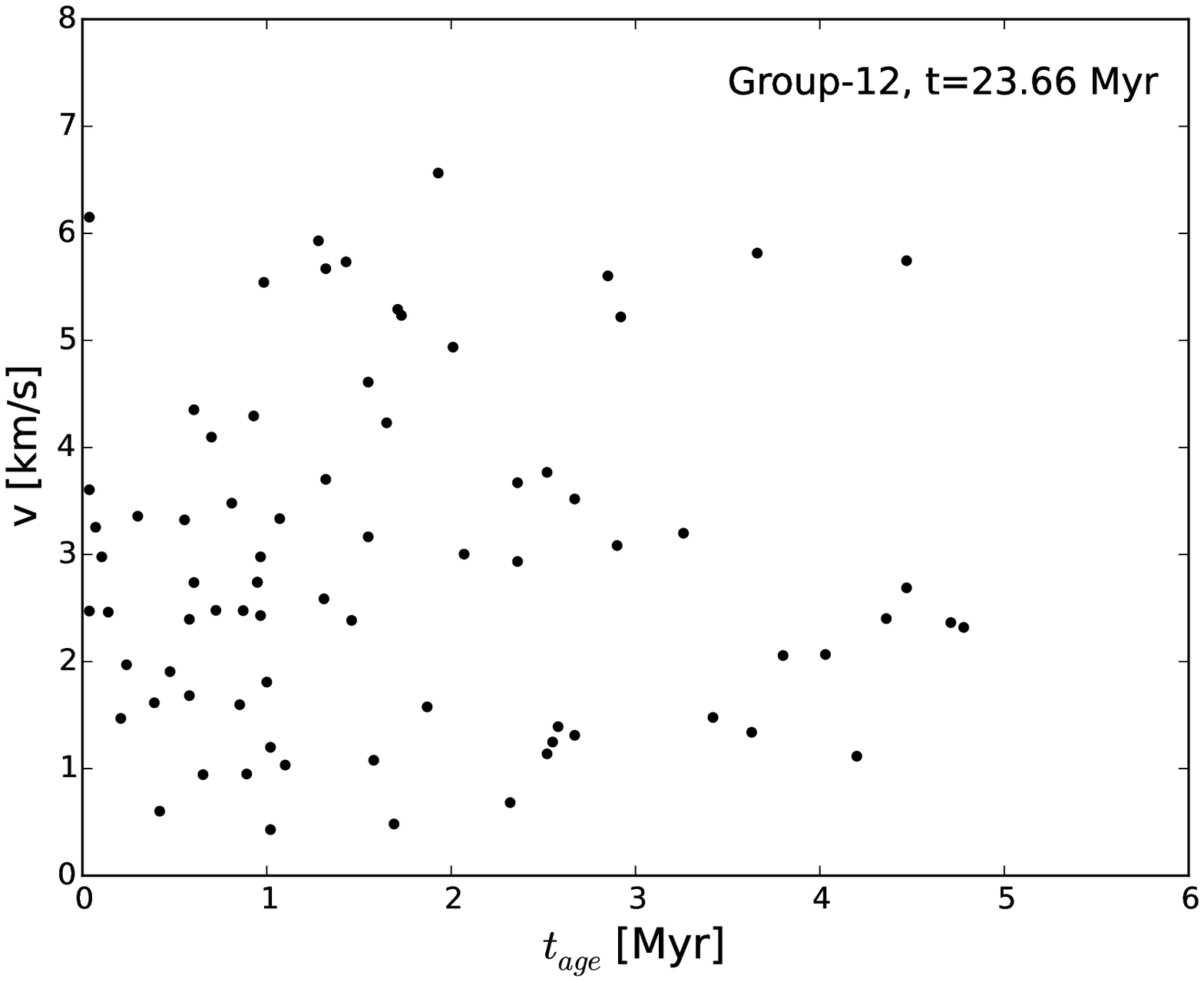}
 \caption{Stellar
   velocity {\it vs.} age for the cluster at times $t=22.40$ and 23.66
   Myr. In the {\it left} panel, the solid line is a fit to the data
   points, indicating the trend of increasing velocity dispersion with
   larger age. }
\label{fig:v-age}
\end{figure*}

Also, as mentioned in Secs.\ \ref{sec:intro} and \ref{sec:SFR}, another
implication of the GHC is that the SFR increases in each star-forming
site until feedback begins to destroy the gas supply onto it, at which
point the SFR begins to decline there, and perhaps shut off completely.
Using a semi-analytic model for the evolution of collapsing clouds and
their SFR, \citet{ZA+12} showed that this acceleration of the SF at each
star-forming site implies that the age histogram of the site before a
few Myr of its destruction is characterized by a large peak of very
young (age $<1$ Myr) and a small tail of older (up to several Myr)
stars, similarly to the age histograms of embedded clusters in various
nearby MCs \citep{PS99, PS00, PS02}.

\begin{figure*}
 \includegraphics[width=0.95\textwidth]{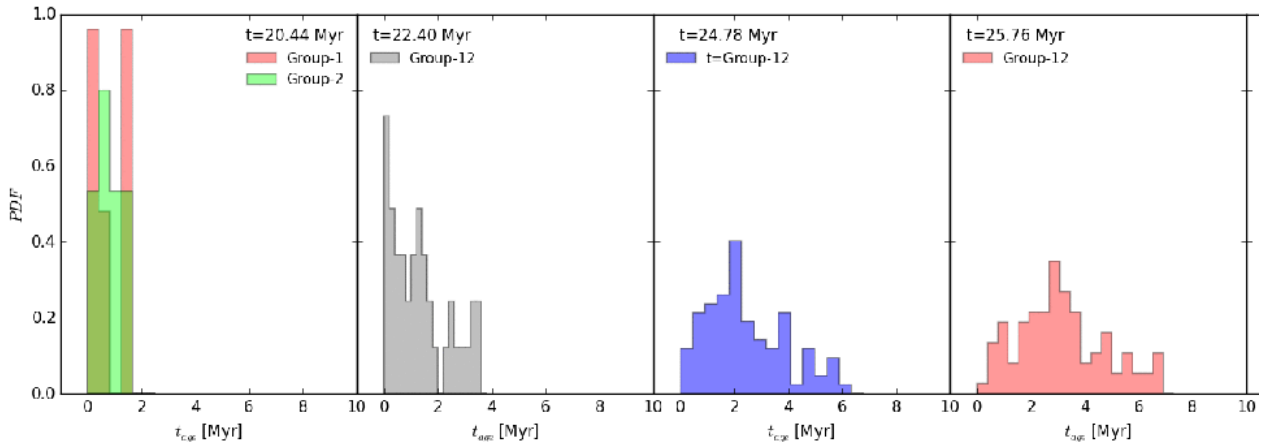} 
\caption{Age histogram of Cluster 2 at various times. }
\label{fig:age_hist}
\end{figure*}

Figure \ref{fig:age_hist} shows the age histogram of the stars in Groups
1 and 2 at various times, as the clumps that embed them merge and
continue to form stars. At $t=20.44$ Myr, the two clumps and their
embedded groups are still separate, and a histogram for each group is
shown. At $t = 22.4$ Myr, the two clumps and their embedded groups have
already merged, and thus a single histogram is shown. It is seen that,
at this time, the most numerous stars are those less than 1 Myr
old. Neverthless, there are also a few stars up to nearly 4 Myr
old. This indicates an acceleration of SF during the first $\sim 4$ Myr
of evolution of Cluster 2. At later times ($t= 24.78$ and $t=25.76$
Myr), the number of young stars decreases, while the most numerous stars
are those formed around $t \approx 22.40$ Myr. This indicates a decline
of the SFR after this time due to the onset of gas dispersal by the
feedback.

\subsubsection{Mass distributions} \label{sec:mass_distr}

\begin{figure*}
 \includegraphics[width=0.49\textwidth]{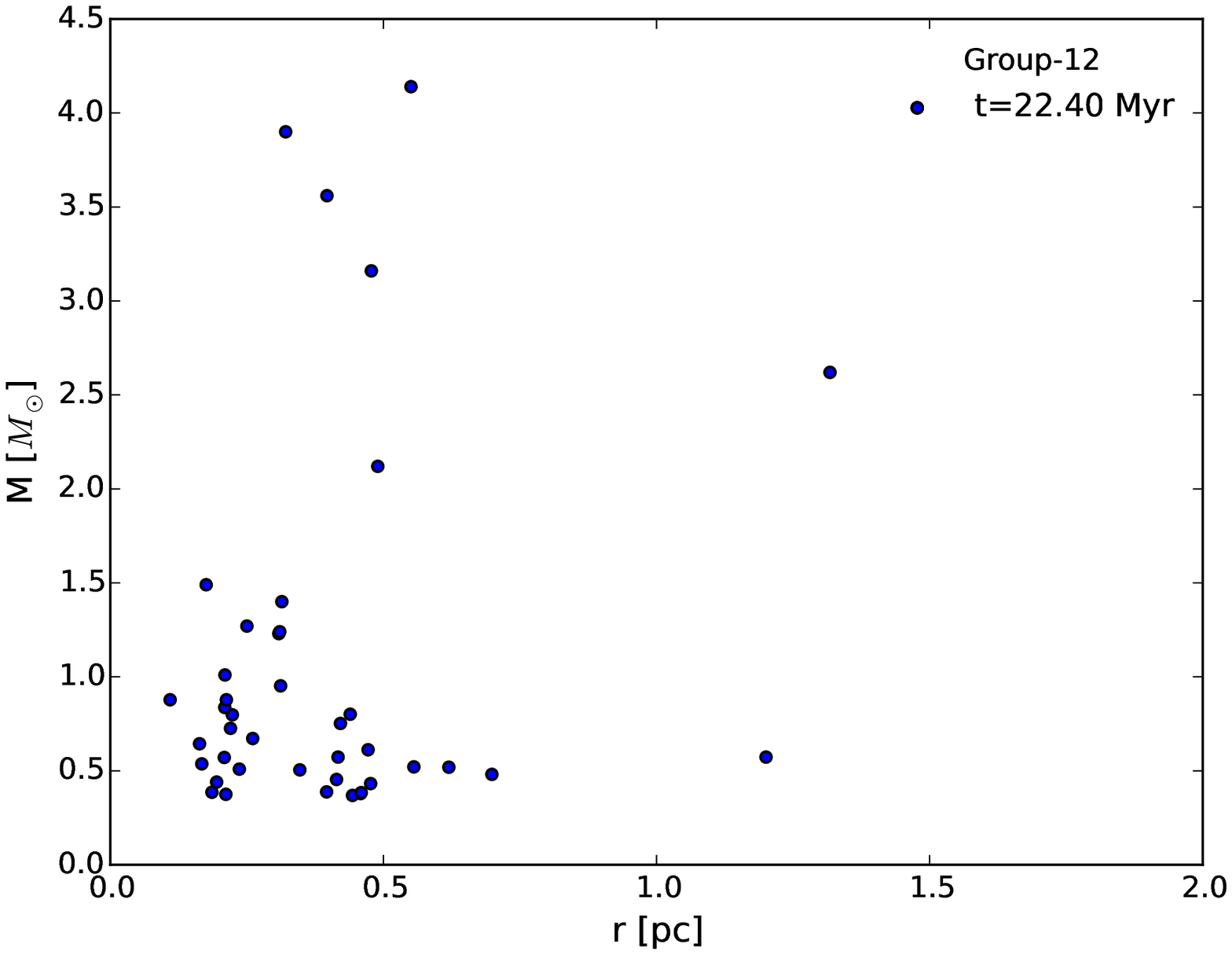} 
 \includegraphics[width=0.49\textwidth]{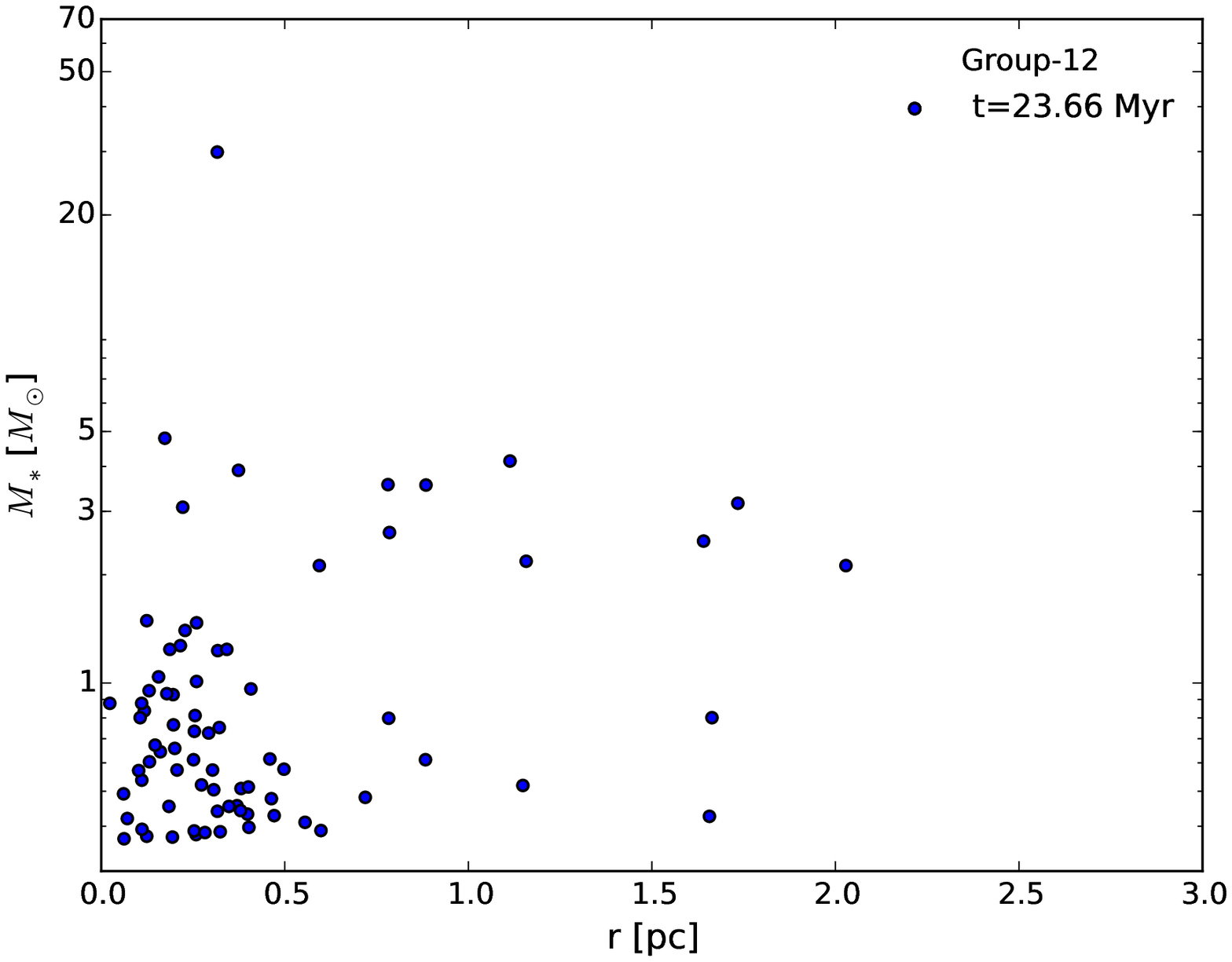} 
\caption{Radial distribution of the stellar particle masses at $t=22.40$
  and 23.66 Myr. Note that the vertical axis is linear in the left
panel, and logarithmic in the right panel.}
\label{fig:mass_profile}
\end{figure*}

Figure \ref{fig:mass_profile} shows the radial distribution of the
stellar particle masses at $t=22.40$ Myr. It can be seen that all but
one of the stars with mass $M >2 \Msun$, and all stars with $M>3
\Msun$ are located within $\sim 0.5$ pc of the computed center of the
cluster. Only one star with $M \approx 2.5 \Msun$ is located at $r
\approx 1.3$ pc from the cluster's center of mass. However, close
inspection of the time sequence outlined by the various images of
Fig.\ \ref{fig:cluster_evol} shows that this is a runaway star formed
in, and ejected from, Group 1 before the merger. This star is
represented by the cyan squared symbol lying on the green solid circle
in the {\it top right} panel [$t = 21.44$ Myr] of Fig.\
\ref{fig:cluster_evol} and on the solid orange circle in the {\it bottom
  left} [ $t = 22.40$ Myr] panel. Thus, this star is just ``flying by''
the cluster. Instead, the other massive
stars in Group 1-2 have formed {\it in situ}, as indicated by the
triangular shape of their symbols in Fig.\ \ref{fig:cluster_evol}.

Also, the other massive stars are represented by symbols of white and
cyan colors, and with triangular shapes, indicating that those stars are
younger ($\tage < 2$ Myr), and formed {\it in situ} in the merged Group
1-2.  Thus, we conclude that, except for the star that is ``passing
by'', the massive stars have formed after the merger of Groups 1 and 2 to
form Group 1-2, thus being younger and tightly clustered near the group
center.

\subsection{Cluster expansion} \label{sec:expansion}

Another feature of the cluster evolution seen in Figs.\
\ref{fig:age_profile} and \ref{fig:mass_profile} is that the cluster
expands as it evolves, increasing its radius from $r \approx 1.4$ pc
at $t=22.40$ Myr to $r \approx 2$ pc at $t = 23.66$ Myr. The reason for
this is not totally clear. It can be due in part to the fact that the
gaseous material is beginning to be expelled from the clump, as can be
seen in Figs.\ \ref{fig:imgs_clus2} and \ref{fig:imgs_clus2b}, and in
part to the fact that the velocity dispersion of the older stars, formed
in the scattered SSLA regions that have fallen into the LSSA one,
corresponds to the large-scale potential well, not that of the local
star-forming site (cf.\ Sec.\ \ref{sec:age_distr}), causing the cluster
to expand to a size representative of this somewhat larger energy.

\subsection{Self-similar cluster structure} \label{sec:hier_clus}

\begin{figure*}
 \includegraphics[width=0.49\textwidth]{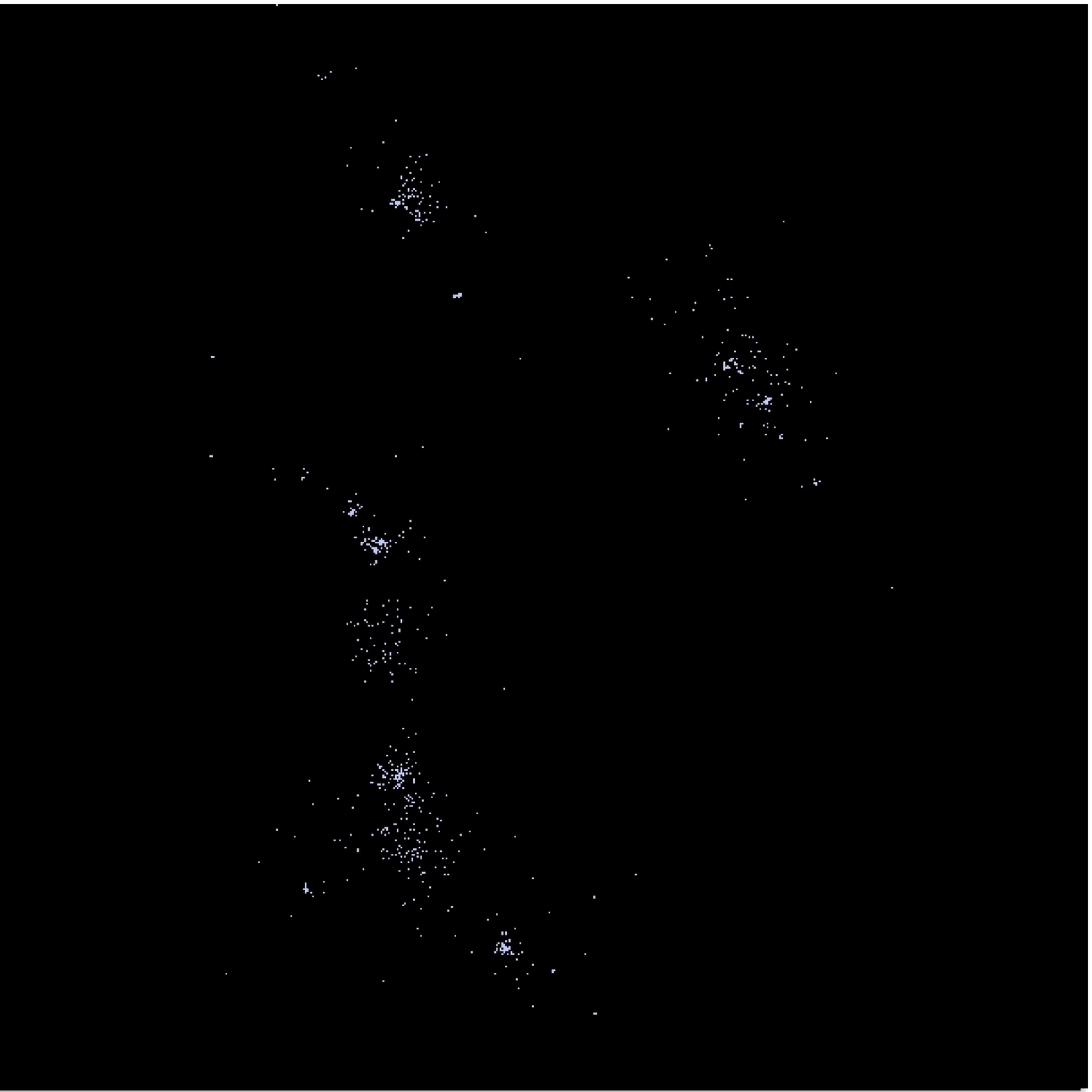}
 \includegraphics[width=0.49\textwidth]{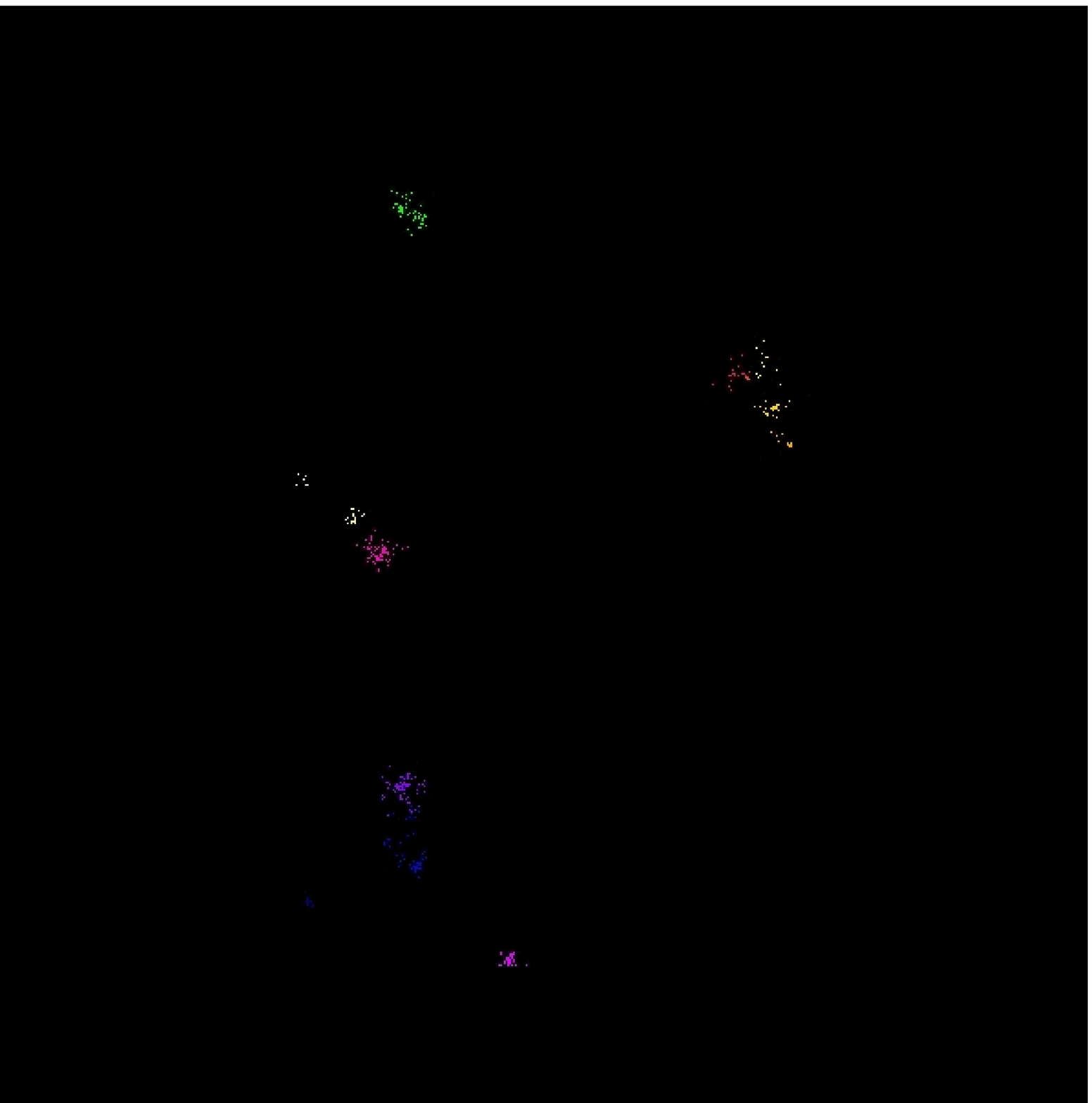}
 \includegraphics[width=0.49\textwidth]{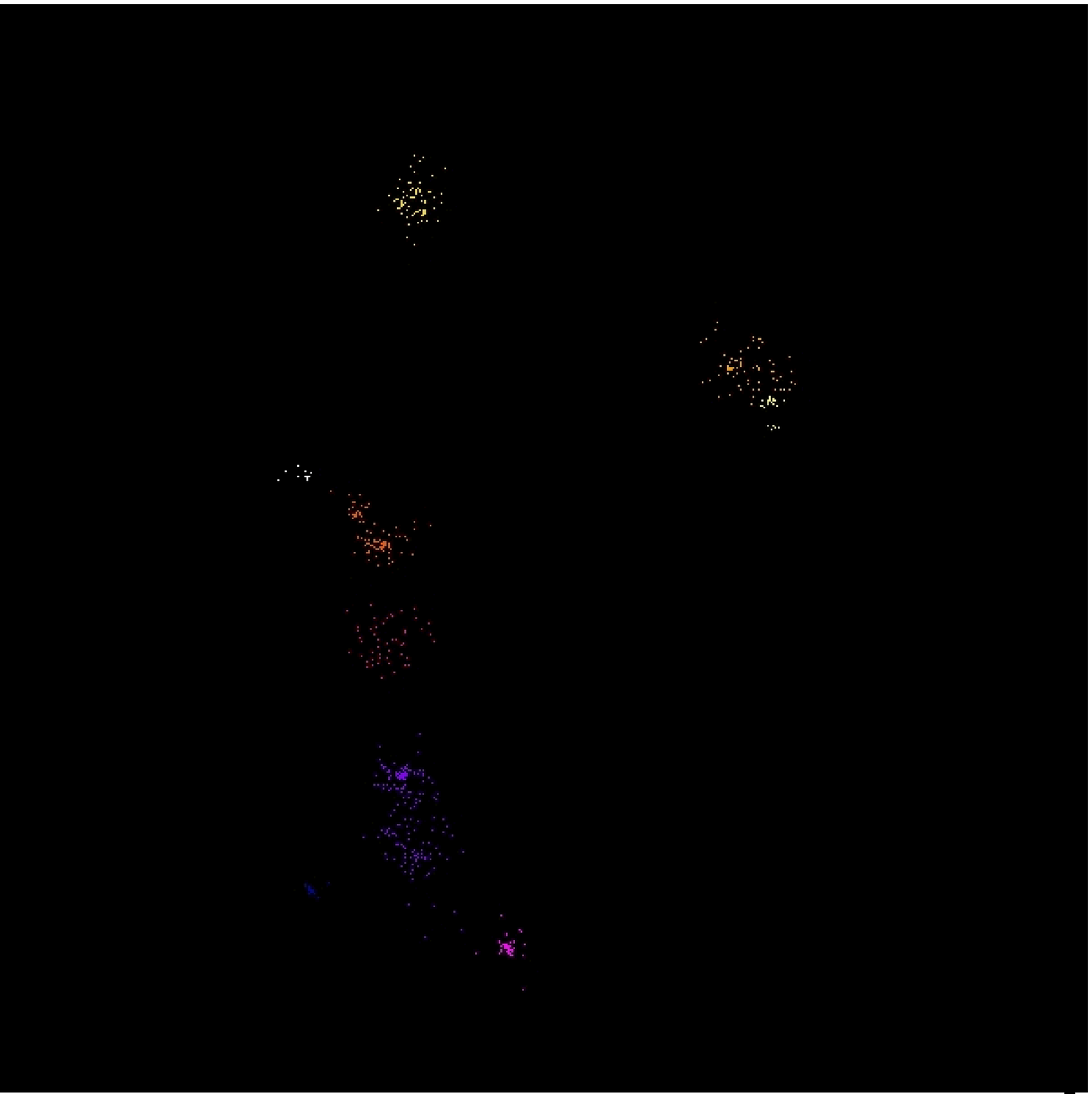}
 \includegraphics[width=0.49\textwidth]{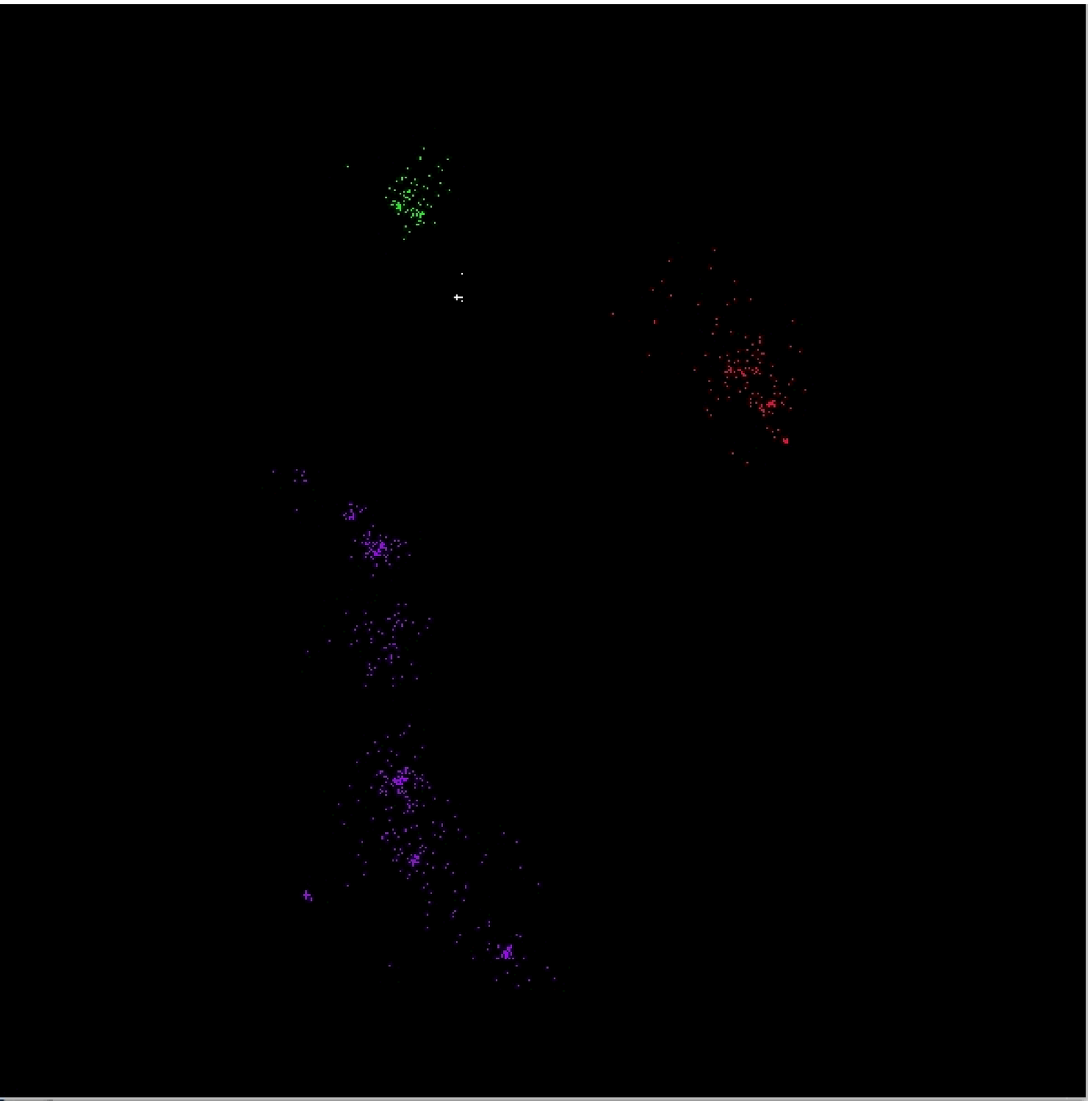}
 \caption{Groups identified by the Friends-of-Friends algorithm in
   Cluster 2 at time $t= 29.96$ Myr varying the
   ``linking parameter'' $\ell$ of the algorithm, which determines the
   distance out to which neghbours are searched. {\it Top left}: All
   stars in the cluster. {\it Top right}: $\ell =0.5$. {\it Bottom
     left}: $\ell =1$. {\it Bottom right}: $\ell = 2$.}
\label{fig:hier_clus}
\end{figure*}

We now discuss the hierarchical structure of the cluster itself,
resulting from the process of GHC of its parent cloud. Figure
\ref{fig:hier_clus} shows the stars belonging to the cluster at time
$t=29.96$ Myr, first
as a whole, and then as identified by a friends-of-friends algorithm
using three different values of the ``linking parameter'' $\ell$, which
determines how far neighbours are searched in order to identify a
group. 
We see that, using a small value of the
parameter ($\ell =0.5$, {\it top right} panel), 10 tight groups are
identified (denoted by the different colors), while using larger values
($\ell = 1$ and 2, {\it bottom left} and {\it bottom right} panels),
respectively 9 and 4 are identified, each group being significantly more
extended. We also note that, at the lowest value, $\ell = 0.5$, a whole
group of moderately scattered stars is left out, which is nevertheless
identified as a group at higher values.

These results show that the structure of the cluster resulting from the
global, hierarchical contraction of the cloud, is inherently ``nested'',
consisting of structures within structures within structures, reflecting
the structure of the parent cloud, which in turn is a
consequence of its multi-scale, hierarchical collapse.

\section{Discussion}
\label{sec:discussion} 

\subsection{Summary of cluster properties resulting from hierarchical
  assembly}
\label{sec:implied_props}

We can summarise the discussion in the previous section as follows. The
underlying assumption for interpreting the emerging properties of the
clusters formed in the scenario of GHC is that first proposed in
\citet{VS+09}, namely that massive stars only form in sites where the
local SFR is high enough that the IMF can be sampled to correspondingly
large stellar masses. Conversely, the most common stars in low-SFR sites
should be those at the peak of the IMF. This is, of course, a fully
probabilistic interpretation of the IMF, and it does not address the
physical mechanisms responsible for the development of the IMF itself.
However, it allows a framework for understanding the assembly of the
clusters.

Also, under the assumption that structures of all scales are collapsing,
as required by the GHC, at a given
density (and therefore, a given free-fall time), more massive clumps are
expected to have higher SFRs, if the SFR is given, to first order, by
the ratio of the clump mass to its free-fall time.

Our simulation mimics these hypotheses, due to the probabilistic scheme
used for the formation of a stellar particle. The longer it takes for a
particle to form, the more massive it will be, since the cell in which
it forms continues to accrete and becomes more massive, without further
refinement, and then the particle forms with half the mass of the cell.
But, as shown in Figure 1 of Paper I, the probability of being able to
wait very long times before forming a stellar particle is very low, and
so the probability of forming of a very massive star is also very low. Thus,
our simulation is expected to generally only form massive stars when the
local SFR is very high, although there can be a few exceptions to this
pattern. 

Finally, in the GHC scenario, the low-mass, small-scale but
large-amplitude (SSLA) star-forming sites should appear earlier in the
evolution of the cloud, because of their shorter free-fall times, while
the high-mass, larger-scale, smaller-amplitude (LSSA) sites should
appear later. Moreover, the low-mass sites should be falling into the
trough of the gravitational potential well of the high-mass ones.
However, low-mass sites continue to appear at all times in the filaments
that feed the massive clumps until the filaments are destroyed by the
feedback. Thus, the low mass stars formed in the low-mass, infalling
clumps share the clumps' infall velocity, and thus should have larger
velocity dispersions than the stars formed in the main, massive clump,
which can have both low or high masses.

The above considerations imply a number of properties of the resulting
clusters: 

\begin{enumerate}

\item The oldest stars tend to have low masses. However, the converse is
  not necessarily true: low-mass stars in general can be either young or
  old. This trend is shown in the {\it top} panel of Fig.\
  \ref{fig:age-mass}, which shows the mass {\it vs.} the age of the
  stars in the cluster at times $t = 22.40$ and 23.66 Myr.

\item The oldest stars have larger velocities (Fig.\ \ref{fig:v-age},
  {\it left} panel), and
  consequently, they may be encountered at larger distances from the
  local star-forming site (Fig.\ \ref{fig:age_profile}).

\item The most massive stars appear later (a few Myr) in the evolution
  of the clump ensemble, when various SSLA sites have merged to form a
  LSSA site.  This is shown in Figure \ref{fig:cum_mass_hist}, which
  gives the cumulative stellar mass histogram for Groups 1 and 2,
  independently and combined, at various times. As time proceeds,
  progressively larger fractions of the stars are seen to be
  massive.

\item The SFR increases during the early stages of the cluster assembly,
  until the time when massive stars begin to form. After this time, the
  SFR decreases, as the ionising feedback from the massive stars begins
  to destroy the clump and its filamentary gas supply. This is seen in
  the evolution of the age histogram, shown in Fig.\ \ref{fig:age_hist}.

\item The oldest stars constitute a minority, since the SFR initially
  increases with time. However, after the appearance of OB stars that
  can partially or completely destroy the local clump and its supplying
  filaments, the SFR decreases again, and thus stars younger than the
  age of the oldest massive stars should be less abundant as well, as
  also shown in Fig.\ \ref{fig:age_hist}.

\end{enumerate}

Note that these properties refer to regions of sizes a few parsecs and
ages of at least a few Myr, and so it specifically does {\it not} refer
to the local and instantaneous stellar production in a single clump, for
which our simulation cannot discern whether massive stars form first or
last, or how long does it take to build a massive star.

Also, note that young star-forming regions may consist of several
low-mass local star-forming sites, and contain no large-scale site yet.
This situation corresponds to times where the first low-mass sites have
not yet merged to form a more massive one.

\subsection{Comparison with observations} \label{sec:compar_obs}

Our numerical results are qualitatively consistent with recent
observational results on the age and mass distributions of young stellar
objects in star-forming regions. In particular, \citet{Getman+14a,
  Getman+14b} have recently reported the existence of age gradients in
the stellar populations of various massive star-forming regions, such
that the youngest stellar objects are located in the obscured regions of
molecular clouds, intermediate-age objects are located in revealed
clusters, and the oldest objects are found in distributed populations.
This is qualitatively consistent with our result that the oldest objects
can be found at large distances from the cluster's center due to their
larger velocity dispersion, in turn due to their formation far from it,
so that their velocities include the component from their infall onto
the central massive clump. Instead, younger objects formed in the clump
lack this component of the velocity, and these are characterised by
somewhat lower velocites. Also, the larger velocity dispersion of the
older objects predicted by the GHC scenario has been observed by
\citet{Mairs+16} in the Orion A cloud.

\citet{Getman+14b} considered several possible scenarios that
could lead to the age segregation they observed. Of these, the one
closest to the mechanism of hierarchical assembly by GHC is their
scenario B, since it invokes the acceleration of SF as proposed by
\citet{PS00}, which is consistent with our scenario \citep{ZA+12}.
However, the similarity between GHC and Scenario B of \citet{Getman+14b}
is not complete, since they did not consider the accretion of both stars
and gas onto the trough of the large-scale potential well.

In addition, \citet{Povich+16} have concluded that the cloud complex
known as the M17 southwest extension (M17 SWex) appears to be lacking
very massive stars in comparison with the number expected from its
estimated SFR. This region consists of several scattered star-forming
cores, separated by distances of several parsecs from each other (M.\
Povich, private communication; see also Busquet et al.\ 2013). This is
consistent with the hierarchical assembly of the cluster in our
simulation, since several isolated, intermediate-mass star-forming
clumps may produce a relatively numerous combined population of YSOs
that may however not yet have formed massive stars, because in the GHC
scenario this only occurs after these smaller regions have merged to
form a massive one. This is again illustrated in Fig.\
\ref{fig:cum_mass_hist}, which shows that the relative abundance of
massive stars increases over time, especially after Groups 1 and 2 have
merged.

We conclude that the scenario of GHC provides an assembly mechanism for
clusters that naturally predicts the observed age segregation and
bottom-heavy IMF of young clusters. In a future contribution we plan to
carry out a detailed quantitative comparison between the clusters in
simulations that sample the IMF to lower masses (see Sec.\
\ref{sec:limits}) and the observed properties of specific clusters.

\subsection{Limitations and impact} \label{sec:limits}

The present study is based on the underlying hypothesis that our
probabilistic stellar particle formation scheme mimics the mechanism
leading to the development of the IMF. That is, we interpret the IMF
strictly as a random sampling process, so that the scarcity of massive
stars is interpreted to imply that the probability of forming one is
very low and given precisely by the IMF. Interestingly, the fact that
our simulation, using a probabilistic criterion for the formation of
stellar particles (cf.\ Sec.\ \ref{sec:SF}), qualitatively reproduces
several observed properties of young clusters, suggests that our
underlying assumption is reasonably realistic. 

Nevertheless, our star formation scheme does have some limitations.
Most importantly, as described in Sec.\ \ref{sec:SF}, the minimmum
stellar-particle mass is $0.39 \Msun$, which is near the peak of the
observed stellar IMF, and thus we are missing stars on
the low-mass side of the IMF. Our imposed IMF is thus a strict power
law, that lacks the turnover at low masses. 

This implies that we are missing a significant number of low-mass stars,
about half of the total number of stars. Their mass, $\sim 26\%$ of the
total mass for a \citet{Kroupa01} IMF, and $\sim 19\%$ for a
\citet{Chabrier03} IMF, is instead deposited into the stars that do form
in the simulation, with masses $0.39 \Msun$ to $61 \Msun$. This may
affect the dynamics of the clusters formed in the simulation, although
not very strongly, since the missing stars have low masses, and are not
expected to affect the dynamics of the higher-mass ones too much,
especially if the tend to form a roughly uniform background. The
fact that the structure of our clusters resembles that of observed young
clusters reinforces this expectation. We nevertheless plan to improve
our star formation scheme in order to produce a more complete
stellar-particle mass spectrum, and with this carry out a more detailed
and quantitative comparison with observations. Meanwhile, we consider
that our study conveys an illustrative first approximation to the
problem of cluster assembly, capturing the essential aspects of the
hierarchical assembly of these objects.

Other limitations of our simulation are that it does not include
magnetic fields nor any form of feedback other than ionisation heating,
in particular supernova (SN) explosions. The neglect of magnetic fields may
not be crucial, since our present understanding is that MCs are
generally magnetically supercritical, meaning that the magnetic field is
in general unable to support them against their own self-gravity
\citep[e.g., ][] {Crutcher12}, and therefore they proceed unimpeded to
global collapse.

The neglect of SN explosions is also not crucial because of two main
reasons. First, it is becoming increasingly accepted that SN explosions
need to occur within the clouds in order to remove significant amounts
of mass from the clouds \citep{IH15}, although on the other hand, only
SNe that explode in low-density environments avoid energy losses that
allow efficient coupling with the gas \citep {Girichidis+16}. Also,
pre-SN feedback is required to allow blastwaves to propagate efficiently
into the ISM \citep{Geen+16}.  Second, and more importantly, our study
suggests that the assembly of the clusters is accomplished mostly during
the early stages (first few Myr) of the clouds' evoloution, during which
few or no SNe are expected to occur, especially if massive stars form
late in the evolution of the cloud, as suggested by the GHC scenario.
So, the ionising feedback included in our simulations is probably the
most relevant form of feedback for regulating the gas flow onto the
sites where the new stars are forming.

Finally, our feedback prescription (Sec.\ \ref{sec:feedback}) is of
course crude, solving the radiative transfer only in an approximate way.
Although it was shown in Paper I that the evolution of an \hii\ region
with this prescription does track the uniform-medium analytic solution
reasonably well, it is possible that our prescription allows evaporation
of regions that should be shaded behind dense clumps, since it does not
take into account the density of the intervening material between the
ionising star and the cell on which the effect of the radiation is being
determined. Thus, our simulations may somewhat overestimate the
evaporation rate of material in filaments containing dense clumps.
However, since the filaments themselves are denser than the background
medium, and their length far exceeds that of isolated clumps and cores,
their column density for situations where the filament is
aligned with the ionising object and with the intervening clump (as in
the case of filaments feeding the main clump where the ionising object
formed) may be comparable to or even larger than that of the intervening
clump. Therefore, the shadowing caused by this clump is likely to be of
secondary importance.  Indeed, Figure 10 in Paper I shows the gradual
evaporation of filaments containing clumps in a way that allows the
formation of ``pillars'', and finally isolated globules within growing
\hii\ regions, similarly to objects like the ``Pillars of Creation'' and
dark globules, respectively. On the other hand, our feedback
prescription allows the simulations to run orders of magnitude faster
than if the radiative transfer is followed in full. We thus conclude
that our prescription provides a sufficiently realistic framework for
the study of the assembly of stellar clusters.

%

\section{Summary and conclusions} \label{sec:conclusions}

In this paper, we have discussed the mechanism of assembly of a cluster
in a numerical simulation of cloud formation and collapse by converging
flows. The assembly proceeds in a hierarchical way as a consequence of
the hierarchical nature of the collapse of its parent cloud
\citep{VS+09}. The hierarchical collapse regime consists of small-scale,
large-amplitude (SSLA) collapses (involving small total masses) within
large-scale, small-amplitude (LSSA) ones (involving large masses, which
are spread out over large scale regions). Collapse at all scales
involves accretion onto the troughs of the potential wells (the collapse
``centers''). The large-scale collapses involve the generation of
filamentary flows that funnel material from the whole large-scale region
down to its collapse center, and the small-scale collapsing regions
``ride'' along these large-scale filamentary flows \citep{GV14}, in a
``conveyor belt'' fashion \citep{Longmore+14}.
Because the small-scale collapses start forming stars earlier (due to
their larger amplitudes, which imply shorter free-fall times), they
often reach the large-scale collapse center already containing stars in
addition to fresh gas.

This mechanism then implies that massive star-forming sites form by the
merging of smaller-scale star-forming regions, which supply both stars
and gas to the larger sites, and thus these larger-scale sites contain a
mixture of locally formed stars and stars that formed at a remote,
smaller-scale site and that are brought to the large-scale site by the
infalling flow. Moreover, the stars that form in the small-scale,
scattered sites have an infall velocity corresponding to the
larger-scale potential well. However, stars formed in the massive center
of the large-scale collapse, form from gas that has probably been
shocked upon its arrival there, and so they form with a somewhat lower
velocity dispersion. 

Finally, this scenario also implies that the SFR must increase with time
during early stages of evolution, and then decrease again as the massive
stars begin to destroy their forming sites \citep{ZA+12}. This
acceleration of the SF process, combined with a face-value
interpretation of the IMF as a probability distribution function of
stellar masses, suggests that massive stars are only expected to form
once the SFR is large enough to sample the IMF to high masses.
Therefore, massive stars should form last in the evolution of the whole
ensemble of star-forming sites, once the low-mass ones have merged to
form a large-mass, high-SFR one.

From these considerations, we have concluded that a cluster that has
formed from the hierarchical collapse of its parent cloud must have the
following characteristics:

\begin{itemize}

\item The oldest stars tend to have low masses, because they formed in
  the low-mass sites early in the evolution of the cloud. However, newer
  stars can be either low-or high mass, as they form in the high-mass
  site formed by the merger of the low-mass ones.

\item The oldest stars tend to have larger velocity dispersions, because
  they are characterized by the infall velocity of the material onto the
  high-mass collapse center. However, this effect tends to be washed out
  by the stellar interaction as the cluster ages.

\item As a consequence, older stars tend to be at larger distances from
  the collapse center (or filament).

\item The most massive stars tend to be younger, because they only form
  when the SFR of the whole region has increased sufficiently to sample
  the high-mass regions of the IMF.

\item As a consequence, the low- and intermediate-mass sites that have
  not concluded the merging process may be defficient in the
  highest-mass stars.

\item Because of the increasing SFR, most stars are young, although a
  small fraction may be as old as several Myr.

\end{itemize}

These features are in good qualitative agreement with observed
properties of young clusters, which seem to have a hierarchical
structure \citep[] [and references therein] {PZ+10}, have an age
gradient, so that the oldest stars tend to be farther from the forming
sites \citep{Getman+14a, Getman+14b} and to have larger velocities
\citep{Mairs+16}. Also, the scenario implies that ensembles of cores in
early phases of evolution will be defficient in massive stars compared
to the expected number form their combined stellar production, as
observed by \citet{Povich+16}. We conclude that the scenario of GHC
allows a clear understanding of the properties of young clusters, in
terms of the multi-scale collapse of their parent clouds.

\section*{Acknowledgments}

E.V.-S.\ is glad to acknowledge enlightening discussions with Matt Povich on
the observed structure of young clusters.  A.G.S.\ was supported by
UC-MEXUS Fellowship.


\end{document}